{}
{}

\newif\ifpublic\publictrue
\newif\ifarxiv\arxivtrue
\newif\ifscipost\scipostfalse

\documentclass[12pt,a4paper]{article}

\usepackage{amsmath,amssymb}
\ifpublic\else\usepackage{showkeys}\fi
\usepackage{graphicx}
\usepackage[bookmarks=true,hyperfigures=true]{hyperref}
\usepackage[usenames, dvipsnames]{color}
\usepackage[nosort]{cite}
\usepackage[bulletsep]{collref}
\usepackage[utf8]{inputenc}

\def\showkeysrefformat#1{{\normalfont\tiny\ttfamily#1}}
\makeatletter
\def\SK@@ref#1>#2\SK@{%
 {\@inlabelfalse\leavevmode\vbox to\z@{%
 \vss\SK@refcolor\rlap{\vrule\raise .75em%
  \hbox{\showkeysrefformat{#2}}}}}}
\makeatother

\usepackage[a4paper,text={160mm,247mm},centering]{geometry}

\RequirePackage{verbatim}

\makeatletter
\newwrite\bibinl@out
\newenvironment{bibtex}[1][\jobname]{%
  \immediate\openout\bibinl@out #1.bib
  \immediate\write\bibinl@out{\@percentchar generated from `\jobname' starting line \the\inputlineno^^J}%
  \def\verbatim@processline{\immediate\write\bibinl@out{\the\verbatim@line}}%
  \@bsphack\let\do\@makeother\dospecials\catcode`\^^M\active\verbatim@start
}%
{\immediate\closeout\bibinl@out\@esphack}
\makeatother

\allowdisplaybreaks[3]

\numberwithin{equation}{section}

\usepackage[font=small,labelfont=bf,width=0.85\textwidth]{caption}

\newcommand{\alignrel}[1][=]{\mathrel{\phantom{#1}}{}}
\makeatletter
\newcommand{\eqsep}{\@ifstar{\hspace{1em}}{\hspace{2em minus 1em}}}
\newcommand{\@eqjoin}[2]{\hspace{#1}#2\hspace{#1}}
\newcommand{\eqjoin}{\@ifstar{\@eqjoin{1em}}{\@eqjoin{2em minus 1em}}}
\makeatother

\renewcommand{\minalignsep}{2em}

\makeatletter
\expandafter\def\expandafter\calc@shift@align\expandafter
  {\expandafter\hook@align@sep\calc@shift@align}
\def\hook@align@sep{\ifnum\xatlevel@=\@ne
  \dimen@\displaywidth
  \advance\dimen@-\totwidth@
  \@tempcntb\maxfields@
  \divide\@tempcntb\tw@
  \@tempcnta\@tempcntb
  \advance\@tempcntb\m@ne
  \global\eqnshift@\dimen@
  \global\alignsep@\minalignsep\relax
  \global\advance\eqnshift@-\@tempcntb\alignsep@
  \global\divide\eqnshift@\tw@
  \ifdim\eqnshift@<\z@
   \global\eqnshift@\z@
  \fi
 \fi}
\makeatother

\makeatletter
\expandafter\def\expandafter\align\expandafter
  {\expandafter\hook@align@\align}
\expandafter\def\expandafter\math@cr@@@align\expandafter
  {\expandafter\hook@align@cr\math@cr@@@align}
\expandafter\def\expandafter\endalign\expandafter
  {\expandafter\hook@align@end\endalign}
\expandafter\def\expandafter\measure@\expandafter#\expandafter1\expandafter
  {\measure@{#1}\hook@align@}
\def\hook@align@{\gdef\hook@align@end{\donumber}%
  \gdef\hook@align@cr{\nonumber}}
\def\hook@align@cr{}
\def\hook@align@end{}

\newcommand{\donumber}{\gdef\hook@align@cr{\gdef\hook@align@cr{\nonumber}}}%
\newcommand{\numberhere}{\donumber\gdef\hook@align@end{}}%
\makeatother

\ifscipost
\usepackage[bitstream-charter]{mathdesign}
\fi

\usepackage{mathfixs}
\ProvideMathFix{autobold,greekcaps,frac,root}
\ProvideMathFix{multskip}

\RequirePackage[extdef]{delimset}
\newcommand{\state}{\ket}

\ProvideMathFix{vfrac,rfrac}
\newcommand{\iunit}{\mathring{\imath}}
\newcommand{\eunit}{\mathrm{e}}
\newcommand{\half}{\rfrac{1}{2}}
\newcommand{\ihalf}{\rfrac{\iunit}{2}}

\newcommand{\Real}{\mathbb{R}}
\newcommand{\Complex}{\mathbb{C}}
\newcommand{\Integer}{\mathbb{Z}}

\newcommand{\alg}[1]{\mathfrak{#1}}
\newcommand{\grp}[1]{\mathrm{#1}}
\newcommand{\gen}[1]{\mathrm{#1}{}}

\newcommand{\envalg}{\grp{U}}
\newcommand{\affine}[1]{#1^{(1)}}

\newcommand{\genJ}{\gen{J}}
\newcommand{\genL}{\gen{L}}
\newcommand{\genM}{\gen{M}}
\newcommand{\genP}{\gen{P}}
\newcommand{\genQ}{\gen{Q}}
\newcommand{\genD}{\gen{D}}
\newcommand{\genC}{\gen{C}}
\newcommand{\genODL}[1][]{\genD^0_{\genRL\if&#1&\else,#1\fi}}
\newcommand{\genODP}[1][]{\genD^0_{\genRP\if&#1&\else,#1\fi}}
\newcommand{\genOCL}{\genC^{0}_{\genL}}
\newcommand{\genOCP}{\genC^{0}_{\genP}}
\newcommand{\genRL}{\genL}
\newcommand{\genRP}{\genP}
\newcommand{\genRDL}[1][]{\genD_{\genRL\if&#1&\else,#1\fi}}
\newcommand{\genRDP}[1][]{\genD_{\genRP\if&#1&\else,#1\fi}}
\newcommand{\genRCL}{\genC_{\genRL}}
\newcommand{\genRCP}{\genC_{\genRP}}

\newcommand{\cybe}{\delimpair{[[}{[.],}{]]}} 
\newcommand{\liebr}{\comm}
\newcommand{\cobra}{\delta}

\newcommand{\casJJ}{\genJ^2}

\newcommand{\casMM}{\genM^2}

\newcommand{\casQQ}{\genQ^2}
\newcommand{\casD}{\casJJ_{\alg{d}(2,1;\epsilon)}}

\newcommand{\algleft}{\mathrm{L}}
\newcommand{\algright}{\mathrm{R}}

\newcommand{\mref}{{\hlcolor{red}\bar m}}
\newcommand{\rnorm}{{\hlcolor{red}\nu}}
\newcommand{\rnormL}{\rnorm_{\genL}}
\newcommand{\rnormP}{\rnorm_{\genP}}
\newcommand{\rtwist}{{\hlcolor{red}\xi}}
\newcommand{\rtwistL}{\rtwist_{\genL}}
\newcommand{\rtwistP}{\rtwist_{\genP}}
\newcommand{\redang}{{\hlcolor{red}\alpha}}
\newcommand{\redpar}{{\hlcolor{red}\beta}}
\newcommand{\redtmod}{{\hlcolor{red}h}}
\newcommand{\Sphere}{{\hlcolor{red}\mathrm{S}}}
\newcommand{\AdS}{{\hlcolor{red}\mathrm{AdS}}}
\newcommand{\trigV}{{\hlcolor{red}V}}
\newcommand{\gaugeDLL}{{\hlcolor{red}\lambda}}

\newcommand{\ratlimpar}{{\hlcolor{red}\lambda}}
\newcommand{\trigp}[1]{{\hlcolor{red}{z_*^{#1}}}}
\newcommand{\genredpar}{{\hlcolor{red}\alpha}}

\newcommand{\superN}{\mathcal{N}}

\newcommand{\Order}{\mathcal{O}}
\newcommand{\der}{\mathrm{d}}
\newcommand{\diff}[2][.]{\mathinner{\der#2\if #1.\else^{#1}\fi}}
\newcommand{\deltadist}{\delta}

\DeclareMathOperator{\Span}{span}
\DeclareMathOperator{\ad}{ad}

\def\[#1\]{\begin{equation}#1\end{equation}}

\providecommand{\href}[2]{#2}

\makeatletter
\def\mr@ignsp#1 {\ifx\:#1\@empty\else #1\expandafter\mr@ignsp\fi}%
\newcommand{\multiref}[1]{\begingroup
\xdef\mr@no@sparg{\expandafter\mr@ignsp#1 \: }%
\def\mr@comma{}%
\@for\mr@refs:=\mr@no@sparg\do{\mr@comma\def\mr@comma{,}\ref{\mr@refs}}%
\endgroup}
\renewcommand{\eqref}[1]{(\multiref{#1})}
\makeatother

\makeatletter
\newcommand{\namedref}[2]{\hyperref[#2]{#1~\ref*{#2}}}
\newcommand{\secref}{\@ifstar{\namedref{Section}}{\namedref{Sec.}}}
\newcommand{\appref}{\@ifstar{\namedref{Appendix}}{\namedref{App.}}}
\newcommand{\tabref}{\@ifstar{\namedref{Table}}{\namedref{Tab.}}}
\newcommand{\figref}{\@ifstar{\namedref{Figure}}{\namedref{Fig.}}}
\makeatother

\let\oldbib=\thebibliography
\def\thebibliography{\phantomsection\addcontentsline{toc}{section}{\refname}\oldbib}

\let\oldtoc=\tableofcontents
\def\tableofcontents{\phantomsection\addcontentsline{toc}{section}{\contentsname}\oldtoc}

\providecommand{\hypersetup}[1]{}
\providecommand{\texorpdfstring}[2]{#1}

\hypersetup{plainpages=false}
\hypersetup{pdfpagemode=UseNone}
\hypersetup{bookmarksnumbered=true}
\hypersetup{pdfstartview=FitH}
\hypersetup{colorlinks=false}
\hypersetup{citebordercolor={.5 1 .5}}
\hypersetup{urlbordercolor={.5 1 1}}
\hypersetup{linkbordercolor={1 .7 .7}}

\usepackage{metastr}

\usepackage{ifpdf}
\ifpdf\else\RequirePackage[active]{srcltx}\fi
\RequirePackage{color}

\newcommand{\hlcolor}{\color}
\ifpublic\renewcommand{\hlcolor}[2][]{}\fi

\newcommand{\remark}[2][]{{\normalfont\sffamily\hspace{1ex}%
  \def\emph{\textsl}\def\textbullet{$\bullet$}
  \def\tmparga{#1}%
  \def\tmpargb{NB}\ifx\tmparga\tmpargb\color[rgb]{0,0,0.8}\fi%
  \def\tmpargb{EI}\ifx\tmparga\tmpargb\color[rgb]{0,0.5,0}\fi%
  \def\tmpargb{}\ifx\tmparga\tmpargb\color{red}\fi%
  \def\tmpargb{}\ifx\tmparga\tmpargb\else \textbf{#1:} \fi%
  #2\hspace{1ex}}}

\ifpublic\renewcommand{\remark}[2][]{}\fi

\metaset{title}{Affine Classical Lie Bialgebras\texorpdfstring{\\}{ }%
for AdS/CFT Integrability}
\metaset{author}{Niklas Beisert, Egor Im}

\begin{document}

\pdfbookmark[1]{Title Page}{title}
\thispagestyle{empty}

\begingroup\raggedleft\footnotesize\ttfamily
\par\endgroup

\vspace*{2cm}
\begin{center}%
\begingroup\Large\bfseries\metapick[print]{title}\par\endgroup
\vspace{1cm}

\begingroup\scshape
\metapick[print]{author}
\endgroup
\vspace{5mm}

\textit{Institut für Theoretische Physik,\\
Eidgenössische Technische Hochschule Zürich,\\
Wolfgang-Pauli-Strasse 27, 8093 Zürich, Switzerland}
\vspace{0.1cm}

\begingroup\ttfamily\small
\verb+{+nbeisert,egorim\verb+}+@itp.phys.ethz.ch\par
\endgroup
\vspace{5mm}

\vfill

\textbf{Abstract}\vspace{5mm}

\begin{minipage}{12.7cm}
In this article we continue the classical analysis of the symmetry algebra 
underlying the integrability of the spectrum in the AdS$_5$/CFT$_4$ and in 
the Hubbard model. We extend the construction of the quasi-triangular Lie 
bialgebra $\alg{gl}(2|2)$ by contraction and reduction studied in the earlier work
to the case of the affine algebra $\affine{\alg{sl}(2)} \times \affine{\alg{d}(2,1;\alpha)}$. 
The reduced affine derivation naturally measures 
the deviation of the classical r-matrix from the difference form. 
Moreover, it implements a Lorentz boost symmetry, 
originally suggested to be related to a q-deformed 2D Poincaré algebra. 
We also discuss the classical double construction for 
the bialgebra of interest and comment on the representation of the affine structure.
\end{minipage}

\vspace*{4cm}

\end{center}

\ifpublic
\newpage
\tableofcontents
\fi

\newpage

\section{Introduction}
\label{sec:intro}

Throughout the last two decades there has been significant progress in 
understanding and applying integrability in the context of AdS/CFT correspondence 
\cite{Maldacena:1997re}.  One of the most prominent manifestations of such 
systems is the duality between free strings on the $\AdS^5 \times \Sphere^5$ 
background \cite{Metsaev:1998it} and planar $\superN=4$ SYM gauge theory 
\cite{Brink:1976bc} (see \cite{Beisert:2010jr,Bombardelli:2016rwb} for reviews).

Starting from the realisation that in the gauge theory the dilatation 
operator can be identified with the Hamiltonian 
of an integrable long-range quantum spin chain \cite{Minahan:2002ve, Beisert:2003tq, Beisert:2003yb, Beisert:2003ys, Beisert:2004ry} 
it was understood that in the asymptotic regime the spectrum can be solved via a Bethe ansatz \cite{Beisert:2005fw, Beisert:2006qh}. 
The underlying magnon S-matrix turns out to be 
fixed by $\alg{psu}(2|2)$ symmetry up to an overall phase
\cite{Beisert:2005tm}, which, in turn, 
can be pinned by particular physical self-consistency constraints 
\cite{Janik:2006dc, Hernandez:2006tk, Arutyunov:2006iu, Beisert:2006ib,Beisert:2006ez,Dorey:2007xn}.
In parallel, integrability of the dual worldsheet theory was shown 
\cite{Bena:2003wd, Arutyunov:2004vx} (see \cite{Arutyunov:2009ga} for a review). 
Perturbative quantisation in the light cone gauge \cite{Frolov:2006cc, Arutyunov:2004yx, Arutyunov:2006ak} 
allowed for a perturbative calculation of the worldsheet S-matrix 
\cite{Klose:2006zd, Roiban:2007jf, Klose:2007rz} that was consistent with the 
all-loop prediction from the gauge theory side \cite{Beisert:2005tm}.
Taming the finite-size corrections was further achieved by applying the 
thermodynamic Bethe ansatz to the worldsheet theory \cite{Arutyunov:2007tc, Arutyunov:2009ur, Bombardelli:2009ns, Gromov:2009bc}. 
It was subsequently transformed into the quantum spectral curve \cite{Gromov:2013pga,Gromov:2014caa}, 
which resulted in a plethora of analytical and numerical calculations of the spectrum to very high loop orders 
(see \cite{Levkovich-Maslyuk:2019awk, Gromov:2017blm} for reviews).

However, the algebraic formulation of integrability in this duality still remains obscure at finite coupling.
Typically, the presence of quantum integrability is complemented with 
the existence of a specific type of algebra structure called quantum group 
\cite{Drinfel'd:1988} (see also \cite{Chari:1994pz}). 
For integrable quantum field theories (with asymptotic boundary conditions)
integrability manifests as the factorisation property of the S-matrix. 
From the algebraic point of view, this feature 
corresponds to the fact that the S-matrix is a representation of the universal 
R-matrix of the underlying quasi-triangular quantum algebra. Indeed, for the 
system at hand the recovered S-matrix does factorise and satisfies the 
quantum Yang--Baxter equation. Having the full quantum algebra with a 
universal R-matrix is desirable for several reasons. For instance, the R-matrix 
can be evaluated in arbitrary representations. 
The dressing phase in this case will in principle follow by the choice of the representation,
and we will have full algebraic control over the solutions to the crossing equation.
Moreover, it will give new insights regarding the origin and existence of the quantum spectral curve equations 
and provide the derivation of various result from first principles. 

The extended algebra governing AdS/CFT integrability appears to be of the kind of a Yangian quantum algebra \cite{Drinfel'd:1985}
as was demonstrated on the string \cite{Bena:2003wd} and gauge \cite{Dolan:2004ps} sides
at strong and weak coupling, respectively.
Progress towards a complete algebraic description 
is complicated by the non-standard nature of the underlying algebra at intermediate coupling.
Nevertheless, many pieces of this puzzle have already been identified:
As was mentioned earlier, the quantisation of the string theory requires a gauge fixing. 
This breaks the full supersymmetry algebra $\alg{psu}(2,2|4)$ to a 
subalgebra involving two copies of $\alg{su}(2|2)$ (with identical central charges) \cite{Arutyunov:2006ak}. 
The same effect can be observed in the spin chain picture upon fixing a vacuum state \cite{Beisert:2004ry}. 
This symmetry was enough to fix the fundamental S-matrix up to its overall phase.
Moreover, the invariance of the S-matrix under a Yangian symmetry was demonstrated 
\cite{Beisert:2006fmy}. However, the Hopf algebra structure of the Yangian is 
of a non-standard type due to the length-changing effects \cite{Gomez:2006va, Plefka:2006ze}. 
The deformed Yangian was also important for the higher representations \cite{Matsumoto:2014cka} 
(which appear for example in the scattering of the bound states \cite{Dorey:2006dq, Chen:2006gq, Chen:2006gp}), 
where the S-matrix can be again fixed up to an overall phase 
\cite{deLeeuw:2008dp, Arutyunov:2008zt, Arutyunov:2009mi}. 
The Yangian symmetry 
also appears at the level of the action \cite{Beisert:2017pnr,Beisert:2018zxs} and scattering 
amplitudes \cite{Drummond:2009fd}. 
There was some progress 
in different realisations of the deformed Yangian \cite{Spill:2008tp, Beisert:2014hya, Matsumoto:2022nrk}. 
However, it is clear that the Yangian is not 
the end of the story. It was understood that there exist additional symmetries 
that do not manifest themselves at the level-0, namely, the secret symmetry
\cite{Matsumoto:2007rh, deLeeuw:2012jf} and the Lorentz boost \cite{Young:2007wd, Borsato:2017lpf}.

Additionally, the integrability of the one-dimensional Hubbard model 
\cite{Hubbard:1963, Essler:2005aa} turns out to be relevant to our case. 
Although this condensed matter model in principle has nothing to do with 
the AdS/CFT correspondence, the algebraic basis for the integrability happens to be the same 
\cite{Beisert:2006qh}.
Namely, the underlying quantum algebra is the Yangian based on the centrally extended $\alg{su}(2|2)$ 
and its R-matrix \cite{Shastry:1986bb} is actually equivalent to the worldsheet 
scattering matrix discussed earlier. 

Provided the lack of standard methods available, it is highly non-trivial to 
identify the full quantum algebra from the provided evidence. This is where the 
classical limit comes in handy. For quantum groups the classical limit 
is described by Lie bialgebras. Correspondingly, the universal R-matrix is replaced 
by the classical r-matrix \cite{Moriyama:2007jt, Torrielli:2007mc}, which 
essentially describes the tree level S-matrix of the quantum field theory 
\cite{Klose:2006zd}. The subsequent analysis revealed that the resulting 
bialgebra is a particular deformation of the loop extension of 
$\alg{u}(2|2)$ \cite{Beisert:2007ty}, where the additional $\alg{u}(1)$ 
generator corresponds to the secret symmetry mentioned earlier. Moreover, 
the classical representation of the Lorentz boost can be identified as well \cite{Borsato:2017lpf}. 
Curiously, this bialgebra can be obtained 
by a procedure that we refer to as contraction and reduction \cite{Beisert:2007ty, Beisert:2022vnc}
applied to a semi-simple Lie superalgebra involving the exceptional algebra 
$\alg{d}(2,1;\epsilon)$ \cite{VanderJeugt:1985} as a factor \cite{Matsumoto:2008ww}.
Therefore, the natural question is whether it is possible to lift the homomorphism 
to one of quantum groups. This idea is further motivated by the evidence 
that the contraction and reduction indeed can be applied to q-deformed 
algebras \cite{Beisert:2011wq, Beisert:2016qei, Beisert:2017xqx}.

However, before addressing the quantum case, we would like to examine a further 
possibility to extend the classical Lie bialgebra. 
In this article we continue the study of the classical limit 
\cite{Beisert:2022vnc}. We extend the construction to the case of the 
affine algebras that contain a non-trivial central element $\genC$ and a 
derivation $\genD$. In conventional cases, the affine extension merely adds 
some mostly trivial relations to the overall structure, but the derivation can 
be viewed as incorporating the difference form property of the parametric r-matrix.
Here we will construct the extension of the (supersymmetric) Poincaré loop bialgebra to 
an affine bialgebra. Eventually the derivations will act as an additional symmetry
that explains how the difference form of the r-matrix is violated. Furthermore, 
the phase of the classical r-matrix receives some constraints, since the 
derivation acts directly on the loop parameter of the algebra. 

The derivation can also be identified with the classical limit of the (1+1)-dimensional 
q-deformed Poincaré boost generator \cite{Young:2007wd,Borsato:2017lpf}. 
Here we will argue that the Lorentz boost symmetry might also have a quantum affine origin. 
It would be in alignment with \cite{deLeeuw:2011fr}, where the secret 
symmetry was related to a quantum affine algebra.

On top of that, quantum affine algebras must play a role for q-deformations of 
the models we discuss here. In fact, the one-dimensional Hubbard model 
does admit a q-deformation and the integrable structure was obtained in terms 
of quantum affine algebras \cite{Beisert:2008tw, Beisert:2010kk, Beisert:2011wq}.
The same was observed for q-deformations of AdS/CFT integrability
\cite{Delduc:2013qra,Delduc:2014kha} and the worldsheet S-matrix was matched 
with the Hubbard model results \cite{Seibold:2020ywq,Arutyunov:2013ega}
(at least at tree level). Therefore, we also discuss the aforementioned Lie 
bialgebra homomorphism in the case of the trigonometric bialgebra structure, since 
this gives the classical limit of the quasi-triangular quantum affine algebras.

The structure of this paper is as follows.
In \secref{sec:rationalbosonic} we illustrate the main ingredients of our 
construction first in the simple case of the bosonic part of the (classical) 
 symmetry algebra and its rational r-matrix. We review the 
main steps of the contraction and reduction \cite{Beisert:2022vnc} and highlight 
the new features due to the presence of the affine derivation and central charge. 
We also discuss the classical double construction of the algebra. We comment 
on possible deformations of our construction, which lead to the realisation 
of alternative non-standard quasi-triangular Lie bialgebras. 
Then, in \secref{sec:rationalsusy} we embed our construction in a superalgebra
and comment on the relations to the Lorentz boost symmetry 
\cite{Borsato:2017lpf}. In \secref{sec:trigonometricbosonic} we extend the discussion to
the trigonometric r-matrix, which is relevant to the q-deformed models. Finally, 
in \secref{sec:conclusions} we summarise our work and sketch possible 
prospects of further research.

\section{Simple Rational Case}
\label{sec:rationalbosonic}

In this section we take the bosonic semi-simple Lie algebra $\alg{so}(2,2)$,
which is a subalgebra of $\alg{sl}(2)\times\alg{d}(2,1;\epsilon)$,
to illustrate the essential steps of the construction.
We start with the introduction of the notion of quasi-triangular affine bialgebras 
with the r-matrix of rational type. Then we review the contraction and reduction 
for the loop algebra \cite{Beisert:2007ty,Beisert:2022vnc} and extend the 
construction to the derivation and central element. 
The generalisation to the supersymmetric case is straightforward 
and is discussed in the \hyperref[sec:rationalsusy]{subsequent section}.

\subsection{Affine Bialgebra}

Before diving into the discussion of the contraction and reduction, let us 
introduce the main notions of affine bialgebras for this section.

\paragraph{Affine Algebra.}

For an arbitrary finite-dimensional simple Lie algebra $\alg{a}$ there exists an associated 
untwisted affine Kac--Moody algebra $\alg{g}=\affine{\alg{a}}$. One formulation of the affine 
extension consists in considering the algebra $\alg{a}[u,u^{-1}]$ of Laurent 
polynomials in the formal variable $u$ valued in the Lie algebra $\alg{a}$
and introducing a central element $\genC$ as well as a derivation $\genD$.
Here we consider a slightly more general setting with an infinite tower of 
derivations $\genD_{n}$ forming the Witt algebra. For a quasi-triangular affine bialgebra 
we eventually have to pick only one element from this tower. 
\unskip\footnote{The rational r-matrix requires the derivation $\genD_{-1}$, 
whereas the trigonometric one uses $\genD_{0}$.} 
Nevertheless, we will keep all the derivations wherever it is 
possible in order to have the flexibility to choose a particular derivation when needed. 
Moreover, a larger set of derivations might be useful in order to generalise our construction to 
algebras with multiple punctures (see the discussion in \secref{sec:parametric_form}).
The defining relations are then
\unskip\footnote{Notice that the current choice of basis is mixed real and imaginary. 
In order to obtain a pure real or imaginary basis one may rescale the structure 
constants by a factor of $\iunit$ or multiply the affine generators by factors of 
$\iunit$ and $-\iunit$ respectively.}
\begin{align} \label{eq:generic_algebra}
\comm{\genD_n}{\genD_m} 
&= 
(m-n) \genD_{n+m},
\\
\comm{\genD_m}{\genJ^a_n} 
&= 
n \.\genJ^a_{n+m},
\\
\comm{\genJ^a_m}{\genJ^b_n} 
&= 
\iunit f^{ab}{}_{c} \.\genJ^c_{n+m} + m \delta_{n+m=0}\.c^{ab}\.\genC,
\end{align}
where $f^{ab}{}_c$ are the structure constants of $\alg{a}$ and $c^{ab}$ is the 
matrix of the Killing form.

\paragraph{Bialgebra.}

A Lie bialgebra $\alg{g}$ is defined as a Lie algebra equipped with a linear 
map $\cobra: \alg{g} \to \alg{g}\otimes\alg{g}$ called cobracket, such that $\cobra$ 
induces a Lie algebra structure on the dual space $\alg{g}^*$ via the transpose 
map $\cobra^*: \alg{g}^*\otimes\alg{g}^* \to \alg{g}^*$ w.r.t.\ the bilinear form 
induced by pairings of the dual vectors. 
The cobracket is also required to be a 1-cocycle
which means that for all $\gen{X},\gen{Y} \in \alg{g}$
%
\[
\cobra\brk!{\liebr{\gen{X}}{\gen{Y}}} = 
\liebr!{\gen{X}_1 + \gen{X}_2}{\cobra(\gen{Y})} 
+ \liebr!{\cobra(\gen{X})}{\gen{Y}_1 + \gen{Y}_2}.
\]
Here and in what follows we use the standard notation 
$\gen{X}_1 = \gen{X} \otimes 1$ and $\gen{X}_2 = 1 \otimes \gen{X}$.

Since our main interest is the study of the classical limit of a 
quasi-triangular quantum algebra, in this paper we focus on the quasi-triangular 
Lie bialgebras, whose cobracket is given in terms the classical r-matrix $r$ by the formula
$\cobra(\gen{X}) = \liebr{\gen{X}_1 + \gen{X}_2}{r}$. The r-matrix has to satisfy 
the classical Yang-Baxter equation (CYBE):
\[ \label{eq:cybe}
\cybe{r}{r} := \liebr{r_{12}}{r_{13}} + \liebr{r_{12}}{r_{23}} + \liebr{r_{13}}{r_{23}} = 0,
\]
and its symmetric part $r_{12} + r_{21}$ must be a quadratic invariant of $\alg{g}$.
In the context of affine Lie bialgebras, we are interested in parametric solutions of
the CYBE that depend on an evaluation parameter $u$. Due to \cite{Belavin:1982}, 
the solutions can be classified by the structure of the poles 
(see also \cite{Abedin:2021}). Here we are particularly interested 
in rational and trigonometric solutions relevant to AdS/CFT integrability. 

\paragraph{Rational r-Matrix.}

A simple rational solution to the CYBE for an affine algebra based
on a simple algebra $\alg{a}$ can be expressed as 
\[ \label{eq:general_simple_rmat}
r = \sum_{k = 0}^\infty c_{ab}\. \genJ^a_k \otimes \genJ^b_{-1-k} + \genC \otimes \genD.
\]
Here our starting point is the affine Kac--Moody algebra 
$\affine{\alg{sl}(2)}$. The level-0 algebra is spanned by 3 generators $\genJ^{0,\pm}$,
with the non-trivial structure constants being 
$f^{0 \pm}{}_{\pm} = - f^{\pm 0}{}_{\pm}  = \pm 1$ and $f^{\pm \mp}{}_{0} = \mp 2$.
The rational r-matrix is given by \eqref{eq:general_simple_rmat},
but we also supplement the r-matrix with a twist term, whose importance will be 
apparent later 
\unskip\footnote{In fact, this r-matrix is almost of the most general form; the only 
admissible additional term (up to automorphisms) is at level-1 
$\genJ^0_0 \wedge \genJ^+_1$ \cite{Stolin:1990}, which we exclude from 
consideration in what follows.}
\[ \label{eq:rmat_sl2} 
r_{\alg{sl}(2)}
=
\rnorm\. \sum_{k=0}^\infty c_{ab}\.\genJ^a_k\otimes \genJ^b_{-1-k} 
+ \rtwist\. \genJ^0_0 \wedge \genJ^+_0
+ \rnorm\.\genC\otimes\genD,
\]
where $\rnorm$ and $\rtwist$ are some arbitrary parameters 
\unskip\footnote{Since the r-matrix can always be rescaled, 
we effectively have a one-parameter family of inequivalent r-matrices}
and the matrix $c_{ab}$ has the non-trivial elements $c_{00}=-1$ and 
$c_{\pm\mp}=\half$.

\paragraph{Parametric Form.}

In many cases it is useful to express the above affine algebra relations 
using the loop parameter $u$ in the functional form. This is achieved by 
writing the loop algebra generators in the form of the polynomial algebra explicitly
\[
\genJ_n = u^n \genJ,
\]
and for arbitrary Laurent polynomials $f(u)$ and $g(u)$ from $\Complex[u, u^{-1}]$
the algebra relations in the parametric form can be expressed as
\unskip\footnote{We define the contour around $u=\infty$
as a large circle with negative order,
or in other words, $\oint_{\infty} \diff{u}/u = -2\pi\iunit$.}
\begin{align} 
\comm!{f(u)\genD}{g(u)\genJ^a} 
&= 
f(u) g'(u)\. \genJ^a,
\\
\comm!{f(u)\genJ^a}{g(u)\genJ^b} &= \iunit f(u)g(u) f^{ab}{}_{c}\. \genJ^c
+ \frac{1}{2\pi \iunit}\oint_\infty f(u)\.\der g(u)\. c^{ab} \.\genC.
\end{align}
The rational r-matrix can also be cast to the functional form
\[ \label{eq:rmat_sl2_par}
r_{\alg{sl}(2)}(u_1,u_2)=-\frac{\rnorm\. \casJJ}{u_1-u_2} 
+\rtwist\. \genJ^0 \wedge \genJ^+
+\rnorm\.\genC\otimes\genD,
\]
however, in the presence of the central charge it is important to consistently keep the 
same expansion of the first term in series over $u_1/u_2$. As a bookkeeping device 
we introduce the distribution $\deltadist_{a,b}(z)$ \cite{Beisert:2010kk} such that 
a contour integral
\[
\int_\gamma f(z) \deltadist_{a,b}(z) \diff{z}
\]
picks up $f(z=0)$ for each directed crossing of $\gamma$ through a cut between 
$a$ and $b$. This allows us to consistently fix residues in all expressions via
\[ \label{eq:rmatrixformalseries}
\sum_{k=0}^\infty u_1^{k} u_2^{-k} 
= 
- \frac{u_2}{u_1 - u_2} + 2 \pi \iunit u_2 \deltadist_{0,\infty}\brk*{u_1 - u_2}.
\]

In evaluation representations,
the affine generators are represented as
\[
\genC\state{u}=0,
\eqsep
\genD_n\state{u}=-u^{n+1} \frac{\partial}{\partial u} \state{u}.
\]
Therefore, the introduction of the central charge
has no impact on the representations discussed here,
but there are interesting applications of affine algebras with central charges.

\subsection{Contraction}

The first step of our construction is the contraction of the square algebra 
$\alg{sl}(2)_1 \times \alg{sl}(2)_2 \simeq \alg{so}(2,2)$ spanned by 
the generators $\genM^a_i \in \alg{sl}(2)_i$, see \cite{Beisert:2022vnc}.
The contraction relations read
\[ \label{eq:contraction}
\genL^a = \genM^a_1+\genM^a_2, 
\eqsep 
\genP^a = \epsilon\mref\genM_1^a,
\]
which give a map to the 3D Poincaré algebra $\alg{iso}(2,1)$ in the limit $\epsilon \to 0$.
Generalisation to the loop algebra is straightforward.

\paragraph{Affine Contraction.}

The contraction of two affine algebras based on $\alg{sl}(2)$
works much as for the contraction of the loop algebras presented in \cite{Beisert:2022vnc}.
This becomes apparent if one recalls that the affine extension amounts to 
adding a central element and an automorphism to a loop algebra. Thus, the 
extension does not interfere with the ``internal'' structure of the 
loop algebra and it commutes with the contraction.
Curiously, the resulting contraction is not merely
the affine extension of $\alg{iso}(2,1)$,
but there are two sets of affine generators.
In order to obtain these,
the contraction limit is performed as before, see \eqref{eq:contraction} and \cite{Beisert:2022vnc},
with the additional change of basis relations
\unskip\footnote{Here we reserve the symbols $\genRDL$ and $\genRDP$ for the subsequent reduction 
(see \secref{sec:rationalbosonic_reduction}), which will require a redefinition 
of $\genODL$ and $\genODP$.}
\begin{align} \label{eq:contraction_affine}
\genOCL&=\genC_1+\genC_2,
&
\genOCP&= \epsilon \mref \genC_1,
\\
\genODL[n]&=\genD_{1,n}+\genD_{2,n},
&
\genODP[n]&= \epsilon \mref \genD_{1,n}.
\end{align}
The resulting relations of affine Poincaré generators can be written as
\begin{align} \label{eq:alg_iso21}
\comm{\genODL[m]}{\genL^a_n} &= n \genL^a_{n+m},
&
\comm{\genL^a_m}{\genL^b_n} &= \iunit f^{ab}{}_{c} \genL^c_{n+m} + m \delta_{n+m=0}c^{ab}\genOCL,
\\
\comm{\genODL[m]}{\genP^a_n} = \comm{\genODP[m]}{\genL^a_n} &= n \genP^a_{n+m},
&
\comm{\genL^a_m}{\genP^b_n} &= \iunit f^{ab}{}_{c} \genP^c_{n+m} + m \delta_{n+m=0}c^{ab}\genOCP,
\\
\comm{\genODP[m]}{\genP^a_n} &= 0,
&
\comm{\genP^a_m}{\genP^b_n} &= 0,
\end{align}
and the algebra of derivations takes then form
\[ \label{eq:witt_alg_iso21}
\comm{\genODL[m]}{\genODL[n]} = (n - m) \genODL[n+m],
\eqsep 
\comm{\genODL[m]}{\genODP[n]} = (n - m) \genODP[n+m],
\eqsep
\comm{\genODP[m]}{\genODP[n]} = 0.
\]
Now, we observe two pairs of affine charges and derivations $\genOCL$, $\genOCP$
and $\genODL,\genODP$. The derivation $\genODL$ acts as a usual derivation 
in the sense that its action only affects the loop level of the other generator. 
On the contrary, the second derivation $\genODP$ translates the Lorentz generators 
$\genL^a$ to the momentum generators $\genP^a$.
One may view this algebra as the one given in \eqref{eq:generic_algebra}, with 
$\alg{a} = \alg{sl}(2)$,
tensored with polynomials in another parameter $v$. Upon declaring the new  
parameter infinitesimal and keeping all terms 
up to $\Order(v^1)$, one recovers $\genL^a_n$, $\genODL[n]$ and 
$\genOCL$ generators at level 0 (of $v$) and $\genP^a_n$, $\genODP[n]$ 
and $\genOCP$ at level 1.

From the point of view of the evaluation representation, these features 
are implemented by considering two loop parameters, $u$ and $v$, which are 
inherited from the fact that the two copies of $\alg{sl}(2)$ have distinct loop counting parameters. 
This evaluation representation takes the form
\begin{align} \label{eq:evaluation_representation}
\genODL[n]\state{u,v} &= -u^{n+1}\frac{\partial}{\partial u} \state{u,v} 
- (n + 1) u^n v \frac{\partial}{\partial v} \state{u,v},
&
\genODP[n]\state{u,v} &= -u^{n+1}\frac{\partial}{\partial v} \state{u,v},
\\
\genL^a_n \state{u,v} &= u^n \genL^a \state{u,v} + nu^{n-1}v \genP^a \state{u,v},
&
\genP^a_n \state{u,v} &= u^n \genP^a \state{u,v}.
\end{align}
Evaluation representations in general have vanishing central charges
$\genOCL\simeq\genOCP\simeq 0$. The particular representation 
that will be interesting to us is the field representation of the underlying 
3D Poincaré algebra
\begin{align}
\label{eq:momrep}
\genL^0\state{p,\phi}_{m,s}
&=\brk*{\iunit \frac{\partial}{\partial\phi}+s}\state{p,\phi}_{m,s},
\\
\genL^\pm\state{p,\phi}_{m,s}
&=\eunit^{\pm\iunit\phi}\brk*{
\pm e_m(p)\frac{\partial}{\partial p} 
+\iunit\frac{e_m(p)}{p}\frac{\partial}{\partial\phi}
+\frac{sp}{e_m(p)+m}
}\state{p,\phi}_{m,s},
\\
\genP^0\state{p,\phi}_{m,s}
&=e_m(p)\state{p,\phi}_{m,s},
\\
\genP^\pm \state{p,\phi}_{m,s}
&=\eunit^{\pm\iunit\phi}p\state{p,\phi}_{m,s},
\end{align}
where $s$ and $m$ are spin and mass of the representation. Obtaining it as 
a contraction limit was discussed in \cite{Beisert:2022vnc}.

\paragraph{Coalgebra Contraction.}

Contraction of the coalgebra is straightforward. One has to explicitly perform the 
change of basis \eqref{eq:contraction,eq:contraction_affine} 
(for finite $\epsilon$) in the r-matrix of 
$\alg{so}(2,2) \simeq \alg{sl}(2)_1 \times \alg{sl}(2)_2$, 
which is simply a sum of two copies of the $\alg{sl}(2)$ r-matrix \eqref{eq:rmat_sl2}. 
However, we also dress the r-matrix 
with the twist term
\[ \label{eq:so22_r_matrix}
r_{\alg{so}(2,2)} 
= 
r_{\alg{sl}(2),1} + r_{\alg{sl}(2),2}
+ \rtwist_{12}\. \genM^0_1 \wedge \genM^+_2,
\]
that will keep the r-matrix within the reduced subalgebra 
(see \secref{sec:rationalbosonic_reduction}) after the contraction. 
The presence of the twist imposes additional constraints on the parameters 
$\rtwist_{1,2}$ in \eqref{eq:rmat_sl2} from the CYBE. We satisfy those by fixing 
$\rtwist_{1} = 0$. In order to take the contraction limit $\epsilon \to 0$
one has to eliminate possible singularities due to negative powers of $\epsilon$
in the r-matrix \eqref{eq:so22_r_matrix}. This is achieved by tuning the parameters 
of the r-matrix as functions of $\epsilon$ (up to $\Order(\epsilon^3)$ terms):
\[ \label{eq:contraction_coefficients}
\rnorm_{1,2}= \pm \rnormL \epsilon \mref + \half\rnormP \epsilon^2 \mref^2,
\eqsep
\rtwist_2 = - \rtwistL \mref \epsilon + \half\rtwistP \mref^2 \epsilon^2,
\eqsep
\rtwist_{12} = - \rtwistL \mref \epsilon - \half \rtwistP \mref^2 \epsilon^2,  
\]
and we obtain the following r-matrix of the affine Poincaré algebra 
\unskip\footnote{As long as the rational r-matrix is concerned, we denote the derivations
at level $-1$ by $\genODL,\genODP$}
\begin{align}
\label{eq:rmat_poincare}
r_{\alg{iso}(2,1)} 
&=
\rnormL \sum_{n=0}^\infty  c_{ab}
\brk!{\genL^a_n\otimes\genP^b_{-n-1} + \genP^a_n\otimes\genL^b_{-n-1}}
+ \rtwistL\. \genL^0 \wedge \genP^+ 
\\
&\alignrel
+ \rnormP \sum_{n=0}^\infty c_{ab}\. \genP^a_n\otimes\genP^b_{-n-1}
+ \rtwistP\. \genP^0 \wedge \genP^+
\\
&\alignrel
+\rnormL\. \genOCP\otimes\genODL
+\rnormL\. \genOCL\otimes\genODP
+\rnormP\. \genOCP\otimes\genODP.
\end{align}
%

\subsection{Reduction}
\label{sec:rationalbosonic_reduction}

In the following, we discuss the reduction described in \cite{Beisert:2022vnc}
when applied to the extension of the affine algebra in the rational case.
We will see that the derivation $\genODL$ needs to be dressed
by the Lorentz generators $\genL^a$,
while the ideal of momentum generators $\genP^a$
needs to be dressed by the central charge $\genRCP$.

\paragraph{Reduced Derivation.}

The reduction first restricts to a sub-algebra of
the Lorentz algebra $\alg{sl}(2)$
spanned by
\[ \label{eq:reduction_ideal}
\genRL_n:=\redpar^{-1} \genL^0_{n+1} - \half\eunit^{-\iunit \redang}\genL^+_n - \half\eunit^{+\iunit \redang}\genL^-_n.
\]
%
As the generators $\genRL_n$ are composed from different loop levels of the $\genL^a_n$,
the plain derivation $\genODL$ does not preserve the form of $\genRL_n$
\[ \label{eq:red_plain_der}
\comm{\genODL[m]}{\genRL_n}
=
n\genRL_{m+n}
+\redpar^{-1}\genL^0_{m+n+1},
\]
%
which has some residual dependency on the generator $\genL^0$
not belonging to the sub-algebra.
This term can be eliminated by adjoining $\genODL[m]$ with some
combination of the Lorentz generators $\genL^\pm_n$
obeying the algebra relations
\[
\comm!{\genL^\pm_m}{\redpar u^{-1}\genRL_n}
=
\mp\genL^\pm_{n+m}
\pm\eunit^{\pm\iunit \redang}\redpar \genL^0_{n+m-1}.
\]
%
The generator $\genL^0$ is singled out on the right-hand side by the combination
\begin{align} \label{eq:der_adj_terms}
\comm!{\half\eunit^{-\iunit \redang}\genL^+_m-\half\eunit^{+\iunit \redang}\genL^-_m}{\genRL_n}
&=
\genL^0_{n+m}
-\redpar^{-1} \brk!{\half\eunit^{-\iunit \redang}\genL^+_{n+m+1}+\half\eunit^{+\iunit \redang}\genL^-_{n+m+1}}
\\
&=
\genL^0_{n+m}
-\redpar^{-2} \genL^0_{n+m+2}
+\redpar^{-1} \genRL_{n+m+1}
.
\end{align}
%
%
The form of the bracket \eqref{eq:der_adj_terms} is suggestive: 
although we cannot eliminate the term $\genL^0_{n + m + 1}$ in \eqref{eq:red_plain_der}
right away by adding the above combination of the 
$\genL^{\pm}_{m-1}$ scaled by $\redpar$ to the plain derivation $\genODL[m]$,
we can shift its level by $2$ obtaining 
\[
\liebr!{\genODL[m] + 
\redpar \brk{\half\eunit^{-\iunit \redang}\genL^+_{m-1}
-\half\eunit^{+\iunit \redang}\genL^-_{m-1}}}
{\genRL_n} = n \genRL_{m+n} + \genRL_{m+n} + \redpar \genL^0_{m+n-1}.
\]
Therefore, we can add an infinite series to the derivation in order to shift the 
term away completely in the form of a telescoping sum.
Thus we define the adjusted derivation generator $\genRDL$
such that its algebra with $\genL$ closes
\unskip\footnote{One could shift the derivation with an infinite series of positive levels.
In that case, the resulting expressions are qualitatively the same.}
\[ \label{eq:reduction_D}
\genRDL[n]
:=
\genODL[n]
+\sum_{k=0}^\infty \redpar^{2k+1}\brk!{\half\eunit^{-\iunit\redang}\genL^+_{n-2k-1}
-\half\eunit^{+\iunit\redang}\genL^-_{n-2k-1}} + \mu_n \genOCL,
\]
%
%
%
where we also add a term proportional to the central charge $\genOCL$, which is 
mostly inconsequential. 
Conversely, the generator $\genODP$ has proper Lie brackets
in the sub-algebra 
without the need for adjustments. Nevertheless we add a central charge to the 
reduced derivation
\[
\genRDP[n] := \genODP[n] - \mu_n \genOCP
\]
and fix the coefficients $\mu_n$ to be
\[
\mu_n = - \rfrac{1}{8} \delta_{n\ge0}n\brk!{\redpar^n+(-\redpar)^n}  
\]
so that the algebra of derivations resembles that of the 3D Poincaré algebra \eqref{eq:witt_alg_iso21}. 
Finally, we identify the central charges 
before and after the reduction without any modifications 
\[
\genRCP = \genOCP,
\eqsep
\genRCL = \genOCL.    
\]
%

\paragraph{Reduced Centre.}

The choice of the reduced Lorentz generators $\genRL_n$ produces an ideal 
in the resulting algebra. 
In \cite{Beisert:2022vnc} we already derived the algebra resulting 
from dividing out this ideal in the absence of the affine extension.
With the affine extension 
the ideal of momentum generators may also involve the affine central charge
\[ \label{eq:red_ideal}
\gen{I}^\pm_n :=\genP^\pm_n-\eunit^{\pm\iunit\redang}\redpar\genP^0_{n-1}+\eta^\pm_n \genRCP
\]
with some constants $\eta^\pm_n$ to be determined.
We fix these constants by considering the algebra relations between $\genRL_m$
and $\gen{I}^\pm_n$
\begin{align}
\comm{\genRL_m}{\gen{I}^\pm_n}&=
\eunit^{\pm\iunit\redang}\brk!{
\pm\eunit^{\mp\iunit\redang}\redpar^{-1}\gen{I}^\pm_{m+n+1}
-\half\redpar\eunit^{-\iunit\redang}\gen{I}^+_{m+n-1}
+\half\redpar\eunit^{+\iunit\redang}\gen{I}^-_{m+n-1}
}
\numberhere
\\
&\alignrel{}
+\eunit^{\pm\iunit\redang}\brk!{
\delta_{m+n=0}
\mp\redpar^{-1}\eunit^{\mp\iunit\redang}\eta^\pm_{m+n+1}
+\half\redpar\eunit^{-\iunit\redang}\eta^+_{m+n-1}
-\half\redpar\eunit^{+\iunit \redang}\eta^-_{m+n-1}
}\genRCP,
\end{align}
and between $\genRDL$ and $\gen{I}^\pm_n$
\begin{align}
\comm{\genRDL}{\gen{I}^\pm_n}
&=
n\gen{I}^\pm_{n-1}
+\half\eunit^{\pm\iunit\redang}\sum_{k=0}^\infty \redpar^{2k+2}
\brk!{\eunit^{-\iunit\redang}\gen{I}^+_{n-2k-3}+\eunit^{+\iunit\redang}\gen{I}^-_{n-2k-3}}
\\
&\alignrel{}
-n\brk[s]*{\eta^\pm_{n-1}\mp \eunit^{\pm\iunit\redang}\sum_{k=0}^\infty \redpar^{2k+1} \delta_{n=2k+2}}\genRCP
\\
&\alignrel{}
-\half\eunit^{\pm\iunit\redang}\sum_{k=0}^\infty \redpar^{2k+2}
\brk!{\eunit^{-\iunit\redang}\eta^+_{n-2k-3}+\eunit^{+\iunit\redang}\eta^-_{n-2k-3}} \genRCP
.
\end{align}
All the coefficients of $\genRCP$ must vanish for the
$\gen{I}^\pm_n$ to span an ideal.
These constraints are solved simultaneously by the assignment
\[ \label{eq:reduction_center_coef}
\eta^\pm_n
=
\pm \eunit^{\pm\iunit\redang}\redpar^n
\sum_{k=0}^\infty \delta_{n=2k+1}
=
\pm \half \eunit^{\pm\iunit\redang} \delta_{n\geq 0}\brk!{\redpar^n-(-\redpar)^n}.
\]
The appearance of the second affine central charge $\genRCL$
in the Lie bracket of two $\genRL$'s (cf. \eqref{eq:alg_iso21})
does not impose restrictions
since it does not involve the ideal. Therefore, it persists in the resulting 
expressions ``as is''.

\subsection{Reduced Affine Algebra}
\label{sec:reduced_affine_algebra}

Here let us summarise the reduced affine algebra
and outline some relevant features.

\paragraph{Loop Algebra.}

Altogether, the reduced affine algebra is obtained as follows:
The Lorentz generators appear only in the following combination
\[ \label{eq:reduction_L}
\genRL_n:=\redpar^{-1} \genL^0_{n+1} - \half\eunit^{-\iunit \redang}\genL^+_n - \half\eunit^{+\iunit \redang}\genL^-_n.
\]
Dividing out the ideal spanned by $\gen{I}^\pm_n$
restricts the momentum generators by the following identifications
\unskip\footnote{One might as well define the generators $\genRP_n$
with a different contribution of the central charge $\genRCP$
given by the replacement $\genRP_n\to\genRP_n+\pi_n \genRCP$.}
\[ \label{eq:reduction_P}
\genP^0_n = \redpar^{-1}\genRP_{n+1},
\eqsep
\genP^\pm_n = \eunit^{\pm\iunit\redang}\genRP_n
\mp\half \eunit^{\pm\iunit\redang} \delta_{n\geq 0}\brk!{\redpar^n-(-\redpar)^n}\genRCP.
\]
Finally, the derivations appear in the following combination
\[
\genRDL[n]
:=
\genODL[n]
+\sum_{k=0}^\infty \redpar^{2k+1}\brk!{\half\eunit^{-\iunit\redang}\genL^+_{-2k+n-1}-\half\eunit^{+\iunit\redang}\genL^-_{-2k+n-1}},
\eqsep
\genRDP[n]:=\genODP[n].
\]

The resulting non-trivial Lie brackets with the derivations read
\begin{align} \label{eq:reducedderivations}
\comm{\genRDL[m]}{\genRL_n}
&=
n\genRL_{n+m}+\sum_{k=0}^\infty \redpar^{2k}\genRL_{n+m-2k},
\\
\comm{\genRDL[m]}{\genRP_n}
&=
n\genRP_{n+m}-\sum_{k=0}^\infty \redpar^{2k}\genRP_{n+m-2k},
\\
\comm{\genRDP[m]}{\genRL_n}
&=
\redpar^{-2}(n+1) \genRP_{n+m+2}- n\genRP_{n+m},
\end{align}
and the generators $\genRL$ and $\genRP$ commute up to some
terms involving the central charges
\begin{align} \label{eq:reducedbrackets}
\comm{\genRL_m}{\genRL_n}
&=
-\redpar^{-2}(m+1)\delta_{n+m+2=0}\genRCL
+m\delta_{n+m=0}\genRCL,
\\
\comm{\genRL_m}{\genRP_n}
&=
-m\delta_{m+n=0}\genRCP
-\half\delta_{m+n\geq 0}\brk!{\redpar^{m+n}+(-\redpar)^{m+n}}\genRCP.
\end{align}
\paragraph{Reduced Bialgebra.}

The coalgebra structure can be obtained from the reduction of the r-matrix 
\eqref{eq:rmat_poincare}. This works almost automatically, we only have to fix 
the twist parameter 
\[ \label{eq:reduction_r_matrix_parameter_fix}
\rtwistL = \rnormL \redpar^{-1} \eunit^{- \iunit \redang} ,
\]
to obtain 
\begin{align} \label{eq:rmat_reduced}
r_{\alg{gl}(1)\times\Complex}
&=
-\rnormL\sum_{k=0}^\infty
\brk!{\genRL_k\otimes \genRP_{-k-1} + \genRP_k\otimes \genRL_{-k-1}}
+ \rtwistP \redpar^{-1} \eunit^{\iunit \redang} \genRP_1 \wedge \genRP_0
\\
&\alignrel{}
+\rnormP\sum_{k=0}^\infty
\brk!{\genRP_k \otimes \genRP_{-k-1}-\redpar^{-2}\genRP_{k+1} \otimes \genRP_{-k}}
\\
&\alignrel{}
+\rnormL \genRCP\otimes \genRDL
+\rnormL \genRCL\otimes \genRDP
+\rnormP \genRCP\otimes \genRDP.
\end{align}

Using the obtained r-matrix we calculate the cobrackets 
\begin{align} \label{eq:reduced_cobrackets}
\cobra\brk{\genRDL}
&=
-\rnormL\sum_{k=0}^\infty\sum_{l=0}^\infty
\half\brk!{\redpar^{k+l}+(-\redpar)^{k+l}}\genRL_{-k-1}\wedge\genRP_{-l-1}
\\
&\alignrel{}
-
\rnormP\sum_{k=0}^\infty
\half\brk!{\redpar^k+(-\redpar)^k}\genRP_{0}\wedge\genRP_{-k-2},
\\
\cobra\brk{\genRDP} &= 0,
\\
\cobra\brk{\genRL_n}
&=
-\rnormL\delta_{n\geq 0}
\genRCP\wedge
\brk[s]*{n\genRL_{n-1}
+\sum_{k=0}^{n-1} \half \brk!{\redpar^k+(-\redpar)^k}
\genRL_{n-1-k}
}
\\
&\alignrel{}
+\rnormL\delta_{n\geq 0}
\genRCL\wedge\brk[s]*{n\genRP_{n-1}-\redpar^{-2}(n+1)\genRP_{n+1}}
\\
&\alignrel{}
+\rnormP\delta_{n\geq 0}\genRCP\wedge\brk[s]*{
n \genRP_{n-1}
-\redpar^{-2}(n+1)\genRP_{n+1}
+\half \brk!{\redpar^{n-1}+(-\redpar)^{n-1}}\genRP_0
}
,
\\
\cobra\brk{\genRP_n}&=
-\rnormL\delta_{n\geq 1}
\genRCP\wedge
\brk[s]*{n\genRP_{n-1}
-\sum_{k=0}^{n-1} \half \brk!{\redpar^k+(-\redpar)^k}
\genRP_{n-1-k}
}.
\end{align}
We observe that the cobrackets of the reduced generators $\genRL_n, \genRP_n$ 
are proportional to the central charges, which vanish for the representations of our interest. 
However, the cobracket of the main derivation $\genRDL$ is non-trivial.
Roughly speaking, it measures by how much the classical r-matrix deviates from a difference form,
and thus it should impose a non-trivial constraint of the scalar phase of the quantum R-matrix.

We have pointed out that the resulting affine algebra possesses two derivations 
and two central charges which are owed to the algebra's origin in a direct product 
of two simple algebras. It is not evident whether this second set 
serves any practical purpose. If not, it is possible to reduce the algebra
further by projecting out the central charge $\genRCL$ by setting $\genRCL=0$;
then $\genRDP$ makes no appearance in the bialgebra relations and can be dropped.

\subsection{Phase Degree of Freedom}

So far we have not discussed one possibility to deform the classical r-matrix: 
the phase degree of freedom. Notice that in the absence of the affine extension, 
the r-matrix \eqref{eq:rmat_reduced} can be supplemented with any combination of 
terms of the form $\genRP_n \otimes \genRP_m$ for $m, n \in \Integer$, since the 
generators $\genRP_n$ are central. From the point of view of the coalgebra such 
modification is completely inconsequential. 
However, our initial motivation is to realise the S-matrix of 
the AdS/CFT as a representation of the quantum R-matrix. Upon the quantisation, 
the phase of the classical r-matrix is translated to a scalar prefactor of the 
R-matrix. It is subsequently constrained by the crossing relations \cite{Janik:2006dc, 
Hernandez:2006tk, Arutyunov:2006iu,Beisert:2006ib,Beisert:2006ez,Dorey:2007xn}.
When the affine generators are introduced the momentum generators cease to be 
central, which leads to some algebraic constraints on the phase degree of freedom. 
It is important to verify that these do not exclude the admissible phase 
proportional to $\genRP_0 \wedge \genRP_1$ \cite{Arutyunov:2004vx,Beisert:2007ty} 
and to understand, how the phase can be generated in the contraction and 
reduction.

\paragraph{Phase from Contraction.} 

Without the affine extension one may add infinitely 
many different terms of the form $\genP^a_n \wedge \genP^b_m$ to the r-matrix 
of the 3D Poincaré algebra \eqref{eq:rmat_poincare}. However,
before the contraction we do not have much freedom to add terms to the r-matrix
(the rational r-matrix of $\alg{sl}(2)$ is fixed up to automorphisms
\cite{Stolin:1990}). It turns out that the phases can be generated by deforming 
the contraction relations. Namely, we can always redefine the contraction as 
\[
\genL^a = \genM_1^a + \genM_2^a + 
\epsilon (X_1{}^a{}_b \genM_1^b + X_2{}^a{}_b \genM_2^b),
\eqsep
\genP^a = \epsilon \mref \genM_1^a.
\]
The matrices $X_{1,2}$ do not depend on $\epsilon$. The deformation produces 
correct algebra relations for $\alg{iso}(2,1)$ if the $X_i$ satisfy the equations
\[
f^{c a}{}_d X_i{}^b{}_c + f^{b c}{}_d X_i{}^a{}_c + f^{a b}{}_c X_i{}^c{}_d = 0,
\]
where $f^{a b}{}_c$ are the structure constants of $\alg{sl}(2)$. 
In the parametric form, the matrices might be as well set to depend on the 
loop counting parameter $X_i = X_i(u)$, which will effectively mix different loop level 
generators in the contraction. The deformed contraction relations, though 
preserving the algebra structure, do change the r-matrix and produce infinitely 
many terms quadratic in the momentum at arbitrary loop level.

This derivation of the phase makes the constraints on the phase degree of freedom 
explicit once the affine structure is introduced. We impose the condition that 
the (possibly deformed) derivation after the contraction still acts as a derivation 
(in the contraction limit $\epsilon \to 0$)
\[
\comm!{\genODL}{f(u) \genL^a} \overset{!}{=} f'(u) \genL^a,
\eqsep
\genODL = \genD_{1} + \genD_{2} + \gen{Y}_\genD,
\]
where $\gen{Y}_\genD$ is some element of the affine $\alg{sl}(2) \times \alg{sl}(2)$ algebra
with which we deform the derivation. It turns out that this condition can only be 
satisfied if  
\[
\gen{Y}_\genD \sim \genC_{1,2}, 
\eqsep
X_1'(u) = X_2'(u),  
\]
which restricts appearance of higher loop level phases in the r-matrix 
and, thus, the only admissible phase is at level 0
\[ \label{eq:possible_phases_iso21}
\sim \genP^+ \wedge \genP^-, \genP^0 \wedge \genP^+, \genP^0 \wedge \genP^-.  
\]
The additional twist term $\genL^0\wedge\genP^+$ in \eqref{eq:rmat_poincare} 
introduces an asymmetry between $\genP^+$ and $\genP^-$ generators and results in 
the exclusion of the $\genP^0\wedge\genP^-$ term. In the end this has no consequences,
since the reduction identifies positive and negative directions and the latter 
two phases in \eqref{eq:possible_phases_iso21} become proportional to each other.

\paragraph{Phase from Reduction.}

Although the phase of the affine 3D Poincaré algebra turns out to be 
restricted to the level-0 terms only \eqref{eq:possible_phases_iso21}, the 
reduced affine algebra still admits infinitely many extensions of the r-matrix by 
the terms 
\[\label{eq:generic_phase}
r^0_{\alg{gl}(1)\times\Complex} = \sum_{n,m\in\Integer}f_{n,m} \genRP_n \otimes \genRP_m.
\]
Yet, the affine structure imposes particular constraints on the parameters $f_{n,m}$. 
We define the following combinations of the parameters
\[
F_1(k,m) 
:= 
k f_{k+1, m} - \sum_{n>0}\redpar^{2n} f_{k+1+2n, m},
\eqsep
F_2(k,m) 
:= 
k f_{m, k+1} - \sum_{n>0}\redpar^{2n} f_{m, k+1+2n}.
\]
The requirement that the symmetric part of $r^0_{\alg{gl}(1)\times\Complex}$ is 
a quadratic invariant results in the equations for all $k,m \in \Integer$
\[ \label{eq:phase_eq_sym}
F_{\text{s}}(k,m) := \half \brk!{F_1(k,m) + F_2(k,m)} = 0,
\]
whereas the CYBE \eqref{eq:cybe}
for the deformed $r=r_{\alg{gl}(1)\times\Complex}+r^0_{\alg{gl}(1)\times\Complex}$
implies the equations on the anti-symmetric part 
\[ \label{eq:phase_eq_asym}
\delta_{k\ge0}F_{\text{a}}(k, m) - \delta_{m\ge0}F_{\text{a}}(m, k) = 0,
\]
where
\[
F_{\text{a}}(k, m) := \half\brk!{F_1(k, m) - F_2(k, m)}.  
\]
One may convince oneself that the equations \eqref{eq:phase_eq_sym,eq:phase_eq_asym} 
admit infinitely many solutions. Clearly, they cannot be generated from the 
3-parameter family of the phase degree of freedom \eqref{eq:possible_phases_iso21}.
Therefore, these terms must be attributed to reduced algebra automorphisms that 
change the r-matrix by adding the terms \eqref{eq:generic_phase}. The invariance 
of the algebra structure is reflected by the above equations on $f_{n,m}$. Since 
we do not have this amount of freedom before the reduction, the existence of 
the automorphisms is equivalent to alternative choices of the reduction relations 
that preserve the ideal \eqref{eq:red_ideal}. 
Therefore, matching of the phase at the classical level gives us some constraints 
on the exact form of the reduction to be used. For instance, the specific reduction 
discussed above does allow for the specific phase term $\genRP_0 \wedge \genRP_1$ 
in the r-matrix \eqref{eq:rmat_reduced}, which corresponds to the classical limit 
of the dressing phase.

\subsection{Parametric Form}
\label{sec:parametric_form}

It is often easier to work with affine algebras in the parametric form
rather than in terms of loop levels.
In this section we consider the parametric form of the algebra 
of interest. This approach uncovers some curious structures that are not 
explicit otherwise.

\paragraph{Reduced Algebra.}

In the parametric form we consider the affine algebra as the vector 
space\footnote{Notice a possible abuse of notation: here we consider the parameter 
$u$ as a loop level counting variable, which coincides with the usual 
evaluation representation parameter. However, we also consider a 
representation with two evaluation parameters 
(cf.~\eqref{eq:evaluation_representation}), in which case one has to replace 
$f(u) \genL^a \to f(u) \genL^a + f'(u) v \genP^a$ in all expressions.}
$\Complex \genD \oplus \alg{a} \otimes \Complex[u, u^{-1}] \oplus \Complex \genC$. 
Starting from $\alg{a} = \alg{sl}(2) \times \alg{sl}(2)$, the contraction
can be performed precisely as before. However, when we consider the reduction 
of the resulting affine $\alg{iso}(2,1)$ some subtleties emerge. 

The reduction in the parametric form consists of identifying the loop $\alg{u}(1)$ 
subalgebra of $\alg{sl}(2)$ as 
\[
\genRL 
:= 
\redpar^{-1} u \genL^0 
- \half \eunit^{-\iunit \redang} \genL^+ 
- \half \eunit^{+\iunit \redang} \genL^-.
\]
We have to ensure that after the projection from the loop $\alg{sl}(2)$ to 
$\alg{u}(1)$ (as vector spaces) the algebra relations stay consistent. This 
requires the derivations to be shifted:
\[
\genRDL :=\genODL + \gamma(u) \brk!{\half\eunit^{-\iunit\redang}\genL^+
-\half\eunit^{+\iunit\redang}\genL^-},
\]
with some $\gamma(u) \in \Complex[u,u^{-1}]$. Now, the bracket between the 
adjusted derivation $\genRDL$ and the reduced generator $\genRL$ closes if 
$\gamma(u)$ satisfies
\[ \label{eq:param_reduction_der_adjustment}
\gamma(u) (u^2 - \redpar^2) \genL^0 = \redpar \genL^0. 
\]
At this point it is tempting to assign 
\[ 
\gamma(u) = \frac{\redpar}{u^2 - \redpar^2},  
\]
which, however, violates the assumption $\gamma(u) \in \Complex[u,u^{-1}]$. 
One possible resolution is to extend the ring of polynomials for the loop 
algebra to $\Complex[u, u^{-1}, (u - \redpar)^{-1}, (u + \redpar)^{-1}]$. 
The resulting algebra is the so-called 4-point loop (affine) algebra~\cite{Bremner:1994}. 
The $\alg{sl}(2)$ 4-point algebra (which would be our starting point in the 
generalised setting) admits a 3-dimensional central extension, $\Integer_2$-grading 
and weak triangular decomposition~\cite{Bremner:1995} and particular connections 
to the Onsager algebra are also established~\cite{Hartwig:2007}. However, the 
bialgebra structure in this case is not known (to the authors of this work). 
The additional poles at $u = \pm \redpar$ seem to appear naturally in the 
reduction procedure, thus it would be interesting to pursue this question 
further in order to incorporate them into a bialgebra.

Alternatively, \eqref{eq:param_reduction_der_adjustment} could be resolved 
if we promote the polynomials $\Complex[u, u^{-1}]$ to the formal series 
(or formal distributions) $\Complex[[u, u^{-1}]]$. The solution for $\gamma(u)$ 
is given by 
\[ \label{eq:reduction_D_param}
\gamma(u) 
= 
\sum_{k = 0}^\infty \redpar^{2k + 1} u^{-2k -2} 
\equiv 
\frac{\redpar}{u^2 - \redpar^2} 
+  2 \pi \iunit \redpar  \deltadist_{0, \infty}(\redpar^2  - u^2),
\]
where in the last equality we expressed the distributional term 
\eqref{eq:rmatrixformalseries} explicitly.
\unskip\footnote{Alternatively, one could replace the 
distributional term by 
$- 2 \pi \iunit u \deltadist_{0, \infty}(u^2 - \redpar^2)$, 
which would lead to qualitatively equal results.}
The purpose of the distribution is to remove the poles at $u = \pm \redpar$ and 
set the correct residues for $\gamma(u)$ at $u = 0, \infty$.

Next, we can identify an ideal of the affine subalgebra 
$\alg{u}(1) \ltimes \Real^3$, which is removed by the following identifications
\begin{align}
f(u)\genP^\pm 
&= 
\eunit^{\pm\iunit\redang}f(u)\genRP
\pm\frac{\eunit^{\pm\iunit\redang}\redpar }{2 \pi \iunit} 
\oint_\infty\brk*{\frac{1}{u^2-\redpar^2} 
+ 2 \pi \iunit  \deltadist_{0, \infty}\brk{\redpar^2 - u^2}} 
f(u)\diff{u} \genRCP,
\\
f(u)\genP^0 
&= 
\redpar^{-1}uf(u)\genRP.
\end{align}
This leads to the non-trivial algebraic relations involving the derivations:
\begin{align} \label{eq:parametricDLDPrelations}
\comm!{\genRDL}{f(u)\genRL}
&=
\brk[s]*{f'(u) +\frac{u f(u)}{u^2-\redpar^2} 
+ f(u) 2\pi \iunit u \deltadist_{0,\infty}\brk{\redpar^2 - u^2}}\genRL
,
\\
\comm!{ \genRDL}{f(u)\genRP}
&=
\brk[s]*{f'(u) - \frac{u f(u)}{u^2-\redpar^2} 
- f(u) 2\pi \iunit \redpar^2 u^{-1} \deltadist_{0,\infty}\brk{\redpar^2 - u^2}}\genRP,
\\
\comm!{ \genRDP}{f(u)\genRL}
&=
 \frac{u^2-\redpar^2}{\redpar^2}
\brk[s]*{f'(u)+\frac{uf(u)}{u^2-\redpar^2}}\genRP.
\end{align}
Notice that in the reduction the first bracket also obtains an additional term 
proportional to 
\[
(u^2 - \redpar^2) \deltadist_{0,\infty}\brk*{\redpar^2 - u^2},
\]
which we assign to be zero since for any Laurent polynomial 
$f(u) \in \Complex[u, u^{-1}]$
\[
\oint_C f(u) (u^2 - \redpar^2) \deltadist_{0,\infty}\brk*{\redpar^2 - u^2} 
= 0,
\]
for any contour $C$. 
The algebra between $\genRL$ and $\genRP$ becomes non-trivial
due to contributions from the remaining central charge
\begin{align} \label{eq:parametricLLLPrelations}
\comm!{f(u)\genRL}{g(u)\genRP}
&=
- \frac{1}{2\pi \iunit}\oint_\infty f(u)\bigg[\der g(u)
-\frac{u g(u)}{u^2-\redpar^2}\diff{u}
\\
&\qquad
-2 \pi \iunit \redpar^2 u^{-1} \deltadist_{0, \infty}(\redpar^2 - u^2)g(u) \diff{u} \bigg]\genRCP,
\\
\comm!{f(u)\genRL}{g(u)\genRL}
&=
-\frac{1}{2\pi \iunit}\oint_\infty f(u) \redpar^{-2}
\brk[s]*{(u^2-\redpar^2)\der g(u) + u g(u)\diff{u}} \. \genRCL.
\end{align}
We notice that if it were not for the distributional term, the integrand in the former 
equation would have had additional residues at $u = \pm \redpar$. This implies that 
if we considered the 3-dimensional central extension of the 4-point loop algebra (with the 2 additional central elements sitting at 
point $u = \pm \redpar$), 
the reduction would effectively mix the different central charges. However, as long as we stick 
to the functions from $\Complex[[u,u^{-1}]]$, the additional poles disappear.

\paragraph{r-Matrix and Coalgebra.}

As we mentioned at the beginning \eqref{eq:rmatrixformalseries}, in the parametric 
form the r-matrix is a formal series in two variables 
$\Complex[[u_1^{\pm1}, u_2^{\pm1}]]$. 
Generally, a product of two formal series is ill-defined. However, if we restrict to the formal series of the form 
\[
\sum_{k=0}^\infty \brk*{ \frac{u_i}{u_j}}^k, \eqsep \sum_{k=0}^\infty \brk*{\frac{\redpar^2}{u_i^2}}^k,  
\]
which appear in our case, the products are well defined.
Then, we can write the rational r-matrix in the parametric form:
\begin{align}
r_{\alg{gl}(1)\times\Complex} 
&= 
\brk*{\rnormL \brk{\genRL \otimes \genRP + \genRP \otimes \genRL} 
+ \rnormP \genRP \otimes \genRP \brk{\redpar^{-2} u_1 u_2 - 1}}
\brk[s]*{\frac{1}{u_1 - u_2} - 2 \pi \iunit \deltadist_{0,\infty}\brk*{u_1 - u_2}}
\\
&\alignrel
+\rnormL \genRCP\otimes \genRDL
+\rnormL \genRCL\otimes \genRDP
+\rnormP \genRCP\otimes \genRDP
.
\end{align}

The structure of the coalgebra follows from the algebra together with the r-matrix.
The cobrackets in the parametric form read
\begin{align}
\cobra\brk!{\genRDL}&=
-\rnormL\frac{u_1u_2+\redpar^2}{(u_1^2-\redpar^2)(u_2^2-\redpar^2)}
\brk!{\genRL\otimes\genRP-\genRP\otimes\genRL}
\\
&\alignrel 
+ 2 \pi \iunit \rnormL \frac{u_1 \deltadist_{0, \infty}(\redpar^2 - u_1^2 )- 
u_2 \deltadist_{0, \infty}(\redpar^2 - u_2^2 )}{u_1 - u_2}
\brk!{\genRL\otimes\genRP-\genRP\otimes\genRL}
\\
&\alignrel
-\rnormP\frac{(u_1^2-u_2^2)}{(u_1^2-\redpar^2)(u_2^2-\redpar^2)} 
\genRP \otimes \genRP
\\
&\alignrel
+ 2 \pi \iunit \rnormP (\redpar^2 - u_1 u_2) \frac{
  u_1^{-1} \deltadist_{0, \infty}(\redpar^2 - u_1^2 ) 
+ u_2^{-1} \deltadist_{0, \infty}(\redpar^2 - u_2^2 )}
{u_1-u_2} \genRP \otimes \genRP,
\\
\cobra\brk!{\genRDP}&=0,
\\
\cobra\brk!{f(u)\genRL}
&=
-\rnormL \genRCP\wedge \brk[s]*{f'_+(u)+\frac{f_+(u)-f_+(\redpar)}{2(u-\redpar)}+\frac{f_+(u)-f_+(-\redpar)}{2(u+\redpar)}} \genRL
\\
&\alignrel{}
-\rnormL \genRCL\wedge
\frac{u^2-\redpar^2}{\redpar^2}
\brk[s]*{f'_+(u)+\frac{uf_+(u)}{u^2-\redpar^2}}\genRP
\\
&\alignrel{}
-\rnormP\genRCP\wedge
\frac{u^2-\redpar^2}{\redpar^2}
\brk[s]*{f'_+(u)+\frac{uf_+(u)-\half\redpar f_+(\redpar)+\half\redpar f_+(-\redpar)}{u^2-\redpar^2}
}\genRP
,
\\
\cobra\brk!{f(u)\genRP}
&=
-\rnormL \genRCP\wedge \brk[s]*{f'_+(u)-\frac{f_+(u)-f_+(\redpar)}{2(u-\redpar)}-\frac{f_+(u)-f_+(-\redpar)}{2(u+\redpar)}} \genRP,
\end{align}
where $f_+(u)$ denotes the projection of $f(u) \in \Complex[[u, u^{-1}]]$ on $\Complex[[u]]$.
We again observe that without the distributional terms the cobracket would mix 
the Laurent polynomials $\Complex[u^{\pm1}]$ and $\Complex[(u\pm\redpar)^{\pm1}]$. 
Therefore, it is natural to expect the possibility to extend the 4-point algebra 
to the bialgebra.

\paragraph{Evaluation Representation.}

The representation of the affine $\alg{iso}(2,1)$ algebra is given in 
\eqref{eq:evaluation_representation,eq:momrep} and the space is spanned 
by the states $\state{u,v,p, \phi}_{m,s}$. 
The reduction restricts the states to the subset with $\phi = \redang$ and 
$p = p(u)$:
\[
\state{u,v}_{m,s}:=\state{u,v,p(u),\redang}_{m,s},
\]
where the momentum and energy are dependent on the spectral parameter as  
\[ \label{eq:irrep_reduction}
p(u)=\frac{\redpar m}{\sqrt{u^2-\redpar^2}},
\eqsep
e_m(u)=\frac{mu}{\sqrt{u^2-\redpar^2}},
\eqsep
\redpar\frac{e_m(u)}{p(u)}=u.
\]
The resulting irrep of the reduced loop algebra reads
\begin{align}
\genRL_n\state\state{u,v}_{m,s} 
&=
u^n \frac{sm}{p(u)}\state{u,v}_{m,s} 
+ v u^{n-1} \brk!{(n+1) u^2 \redpar^{-2} - n} p(u) \state\state{u,v}_{m,s} ,
\\
\genRP_n\state\state{u,v}_{m,s}
&=
u^n p(u) \state{u,v}_{m,s}.
\end{align}

Now, we extend the representation to the affine case. 
Clearly, these states have no central charge $\genRCL\simeq\genRCP\simeq0$.
It remains to show that the reduced derivation
\[
\genRDL=\genODL +
\frac{\redpar}{u^2-\redpar^2}\brk!{\half\eunit^{-\iunit \redang}\genL^+
-\half\eunit^{+\iunit \redang}\genL^-}
\]
acts consistently
on these states. The derivation $\genODL$
acts on the original states by a derivative with respect to the spectral parameter
\[
\genODL[-1]\state{u,v,p,\phi}=-\frac{\partial}{\partial u} \state{u,v,p,\phi}.
\]
The additional terms in $\genRDL$ proportional to the Lorentz generators $\genL^\pm$
act on the momentum representation as a derivative with respect to the momentum $p$,
see \eqref{eq:momrep}
\[
\brk!{\half\eunit^{-\iunit\redang}\genL^+-\half\eunit^{+\iunit\redang}\genL^-}\state{u,v,p,\redang}_{m,s}
=e_m(p)\frac{\partial}{\partial p} \state{u,v,p,\redang}_{m,s}.
\]
Relations \eqref{eq:irrep_reduction} imply that
\[
\frac{\redpar}{u^2-\redpar^2}=\frac{p(u)^2}{\redpar m^2},
\eqsep
\frac{\partial p}{\partial u}
=
- \frac{e_m(u) p(u)^2}{\redpar m^2} .
\]
Putting the terms together and using the derivative relationship
\[
\frac{\der}{\der u}\state{u,v}_{m,s}
=
\frac{\partial}{\partial u}\state{u,v,p,\redang}_{m,s}
+
\frac{\partial p}{\partial u}\frac{\partial}{\partial p}\state{u,v,p,\redang}_{m,s},
\]
we find that the derivation $\genRDL$ is represented by a total derivative
\[
\genRDL\state{u,v}_{m,s}=-\frac{\der}{\der u}\state{u,v}_{m,s}.
\]
The second derivation $\genRDP$ acts simply as a partial derivative
on the second spectral parameter $v$
\[
\genRDP\state{u,v}_{m,s}=-\frac{\partial}{\partial v}\state{u,v}_{m,s}.
\]

\subsection{Classical Double}
\label{sec:rat_double}

In the following we will show that the bialgebra of interest 
can be realised as a classical double 
thus putting the novel quasi-triangular bialgebra on firmer ground. 

Let us first recall the notion of the classical double \cite{Chari:1994pz}:
Let $\alg{g}_+$ be a Lie bialgebra with the cobracket 
$\cobra: \alg{g}_+ \to \alg{g}_+ \otimes \alg{g}_+$.
The dual of the cobracket induces a Lie algebra structure on the dual space 
$\alg{g}_- = (\alg{g}_+)^*$ with the Lie bracket 
$\liebr{\cdot}{\cdot}_{\alg{g}_-} = \cobra^*(\cdot \otimes \cdot)$.
Then there exists a quasi-triangular Lie bialgebra structure on the sum 
$\alg{g} = \alg{g}_+ \oplus \alg{g}_-$ such that the inclusion of $\alg{g}_\pm$ 
in $\alg{g}$ is a Lie bialgebra homomorphism. Namely, if we fix a basis $\{\genJ^a\}_a$ 
of $\alg{g}_+$ and its canonically dual $\{ \brk{\genJ^a}^* \}_a$, we can write 
the Lie bracket between elements of $\alg{g_+}$ and $\alg{g_-}$ as
\[
\liebr{\genJ^a}{\brk{\genJ^b}^*}_{\alg{g}} 
= 
 \iunit f^{ca}{}_{b} \brk{\genJ^c}^*
+ \iunit d^{a}{}_{bc} \genJ^c,
\]
where $d^{a}{}_{bc}$ are the structure constants of the coalgebra. The classical r-matrix 
$r \in \alg{g}_+ \otimes (\alg{g}_+)^*$ is given as an identity operator on 
$\alg{g}_+$.

An affine Kac--Moody algebra $\alg{g}$ based on a simple algebra $\alg{a}$ with the 
bialgebra structure induced by the r-matrix of the rational type gives an example of 
such classical double. The corresponding Manin triple is 
\[
\brk{\alg{g}, \alg{g}_+, \alg{g}_-} = 
\brk{\Complex \genD \oplus \alg{a}[u, u^{-1}] \oplus \Complex \genC, 
\alg{a}[u] \oplus \Complex \genC, 
\Complex \genD \oplus \alg{a}[u^{-1}]u^{-1}}.
\]
The $\alg{g}_\pm$ subalgebras are isotropic w.r.t.\ the
non-degenerate symmetric form $\brk[a]{ \cdot, \cdot}$ defined by the non-trivial 
pairings:
\[
\brk[a]{ \gen{J}^a_n, \gen{J}^b_m} 
= \frac{c^{a b}}{\rnorm} \delta_{n+m+1},
\eqsep
\brk[a]{\genD, \genC} = \frac{1}{\rnorm},
\]
where $\genJ^a_n$ is a level-$n$ generator of the loop algebra $\alg{a}[u, u^{-1}]$ 
and $c^{a b}$ is the matrix of the Killing form. The dualisation 
$\brk{\alg{g}_+}^* \simeq \alg{g}_-$ is induced by 
the inner product viewed as an action of a vector on its dual ($n \ge 0$):
\begin{align}
\brk{\genJ^a_n}^* 
&= 
\rnorm c_{ab}  \genJ^b_{-n-1},
\\
\genC^*
&=
\rnorm \genD.
\end{align}
The bialgebra structure on $\alg{g}_-$ is given by the usual Lie brackets for 
the polynomial (in $u^{-1}$) 
algebra with the usual derivation and the cobracket defined for 
$\gen{X} \in \alg{g}_-$
\footnote{$\alg{g}_-$ is not coboundary, nevertheless we formally write 
the cobracket as expressed in terms of the rational r-matrix, since the formal 
computation gives the correct coalgebra relations}
\[
\cobra(\gen{X}) = 
\frac{\liebr{\gen{X}_1}{\casJJ_{12}} + \liebr{\gen{X}_2}{\casJJ_{12}}}{u - v},
\]
where $\casJJ \in \alg{a}\otimes \alg{a}$ is the quadratic invariant.

In our case we set $\alg{a} = \alg{sl}(2)\times\alg{sl}(2)$. However, the r-matrix 
that we consider is not the simple rational, but a twisted one \eqref{eq:so22_r_matrix}.
In order to account for the twist terms in the r-matrix, the dualisation needs to 
be deformed. Namely, as before we consider the algebra of polynomials with two central 
charges $\alg{g}_+ = \alg{a}[u] \oplus \Complex \genC_{1,2}$ and introduce the dualisation
\begin{align}
\brk{\genJ_{i,n}^a}^* 
&= 
\rnorm_i c_{ab} \genJ_{i,-n-1}^b
+ \delta_{n,0} \brk!{ \delta_{i,2} \rtwist_2\brk{\delta^{a,0}  \genJ_{2,0}^+ - \delta^{a,+} \genJ_{2,0}^0} 
+ \rtwist_{12} \brk{\delta_{i,1} \delta^{a,0} \genJ_{2,0}^+ - 
\delta_{i,2} \delta^{a,+} \genJ_{1,0}^0}
},
\\
\genC_i^* 
&= \rnorm_i \genD_i.
\end{align}
One can verify that this dualisation induces a consistent Lie algebra structure 
on $\alg{g}_- = (\alg{g}_+)^*$ and the resulting cobracket is indeed a 1-cocycle.
Thus, the coalgebra structure on both algebras is well-defined.
Moreover, the r-matrix of the classical double is given by the twisted r-matrix 
\eqref{eq:so22_r_matrix}.

Now, we are in the position to apply the contraction procedure to the double 
construction. For finite $\epsilon$ the contraction relations \eqref{eq:contraction_affine}
simply amount to a change of basis, which allows us to obtain the dualisation in terms 
of $\genL^a$ and $\genP^a$ generators. This dualisation becomes singular in the 
contraction limit $\epsilon \to 0$, unless the parameters of the dualisation 
are tuned according to \eqref{eq:contraction_coefficients}, resulting
in the dualisation in the limit
\begin{align}
\brk{ \genL^a_n}^*
&=
\rnormL c_{ab}  \genP^b_{-n-1} 
+ \delta_{n,0} \delta^{a,0} \rtwistL \genP^+_0,
\\
\brk{\genP^a_n}^*
&=
c_{ab} \brk{\rnormL \genL^b_{-n-1}  + \rnormP \genP^b_{-n-1}}
+ \delta_{n,0} \brk{\delta^{a,0} \rtwistP \genP^+_0 - \delta^{a,+} \rtwistP \genP^0_0 
- \delta^{a,+} \rtwistL \genL^0_0 },
\\
\brk!{\genOCL}^* &= \rnormL \genODP,
\\
\brk!{\genOCP}^* &= \rnormL \genODL + \rnormP \genODP.
\end{align} 

Finally, we can perform the reduction on the classical double as well, albeit it 
requires additional care. First, we restrict the contracted bialgebra 
$\alg{iso}(2,1)[u] \oplus \Complex \genC_{\genL,\genP}$ to its subalgebra spanned 
by the momentum directions and the reduced $\genRL_{n\ge0}$ generators 
\eqref{eq:reduction_L}.
The dual of the reduced generators is then
\[
\brk{\genRL_n}^* = \brk{ \redpar \genL^0_{n+1}}^*.
\]
The choice of the reduced generators 
$\genRL_{n \ge 0}$ singles out an ideal in 
the momentum subalgebra:
\[ \label{eq:reduction_ideal_cd}
 \gen{I}^{\pm}_{n>0}
= 
\genP^\pm_n 
- \eunit^{\pm \iunit \redang} \redpar  \genP^0_{n-1} + \eta^\pm_n \genOCP,
\eqsep
 \gen{I}^-_0 = \genP^-_0 - \eunit^{- 2 \iunit \redang} \genP^+_0,
\]
with $\eta^\pm_n$ given in \eqref{eq:reduction_center_coef}. The remaining generators 
are given by 
\[
 \genRP_{n\ge0} = \eunit^{-\iunit \redang} \genP^+_n 
+ \eunit^{-\iunit \redang} \eta^+_n \genOCP.
\]
Notice, that as long as 
here we only consider the non-negative loop levels, the addition of the central 
charge $\genOCP$ is (almost) arbitrary. However, this affects the dualisation 
of the central charge $\genRCP = \genOCP$ in the reduced algebra:
\[
\genRCP^*
=
\brk!{\genOCP}^*
-  \sum_{n = 0}^\infty \eta^+_n \brk*{\genP^+_n + \frac{\rnormP}{\rnormL} \genL^+_n}^* 
+ \eta^-_n \brk*{\genP^-_n + \frac{\rnormP}{\rnormL} \genL^-_n}^*.
\]
The terms $(\genL^\pm_n)^*$ 
do not play a role in restricting the algebra, we add them in order to match the 
final expression with previous sections.
The dual of the reduced $\genRP$ directions are 
\begin{align}
\brk{\genRP_{n>0}}^* 
&= 
\brk{\redpar^{-1} \genP^0_{n-1}
+ \eunit^{\iunit \redang} \genP^+_n
+ \eunit^{- \iunit \redang} \genP^-_n}^*,
\\
\brk{\genRP_0}^* 
&= \brk{ 
\eunit^{\iunit \redang} \genP^+_0
+ \eunit^{- \iunit \redang} \genP^-_0}^*.
\end{align}
As the next step, we have to divide out the ideal \eqref{eq:reduction_ideal_cd}
and perform a dual procedure of modding out an ideal in the dual space such that 
the resulting dual algebra is spanned by the generators 
$\set{\brk{\genRL_{n\ge0}}^*, \brk{\genRP_{n\ge0}}^*, \genRCP^*, \genRCL^* = \brk{\genOCL}^*}$. In fact, identification 
of the reduced generators in the dual space is analogous. We define the reduced 
$\genRL_{n<0}$ generators with the same relation \eqref{eq:reduction_L} and the 
ideal spanned by the same combination of $\genP^a$ generators 
\eqref{eq:reduction_ideal_cd} (without the central charges). 
However, there is one issue with the $\brk{\genP_0}^*$ generator: it does not 
belong to the reduced algebra unless the parameter $\rtwistL$ is tuned to be 
$\rtwistL = \eunit^{-\iunit \redang}\rnormL/\redpar$, which is precisely the 
constraint \eqref{eq:reduction_r_matrix_parameter_fix}. After dividing out 
the ideal from the dual algebra the final form of the dualisation is 
\begin{align} \label{eq:reduced_dualisation}
\brk{\genRL_{n\ge0}}^* 
&= 
- \rnormL \genRP_{-n-1},
\\
\brk{\genRP_{n\ge0}}^* 
&= 
- \rnormL \genRL_{-n-1} 
+ \rnormP \genRP_{-n-1} 
- \delta_{n > 0} \rnormP \redpar^{-2} \genRP_{-n+1}
- \delta_{n = 0} \rtwistP \eunit^{\iunit \redang} \redpar^{-1} \genRP_1,
\\
\genRCL^* &= \rnormL \genRDP,
\\
\genRCP^* &= \rnormL \genRDL + \rnormP \genRDP.
\end{align}
%
%
%
This construction shows that the resulting affine algebra can be realised as 
a classical double based on polynomials valued in a 2-dimensional abelian 
algebra centrally extended by two charges according to \eqref{eq:reducedbrackets} 
with the dualisation given by \eqref{eq:reduced_dualisation}.

\subsection{General Reduction}
\label{sec:gen_reduction}

As we have just seen, the classical bialgebra relevant to the AdS/CFT integrability 
can be obtained as a particular reduction. The choice of the reduction seems to 
be rather arbitrary otherwise, and there could be other possibilities leading to a consistent 
quasi-triangular bialgebra (e.g.\ the classical bialgebra of the q-deformed AdS/CFT 
can be obtained as a trigonometric version of the reduction above, see 
\secref{sec:trigonometricbosonic}). The interesting questions then are when such 
reductions are compatible with the bialgebra structure and what happens to the double 
construction.

\paragraph{General Case.}

At first let us consider a completely generic reduction. We fix an arbitrary vector 
within $\alg{sl}(2)[u,u^{-1}] \subset \alg{iso}(2,1)[u,u^{-1}]$
\[ \label{eq:genRedL}
\genRL = \sum_{s \in \set{0, \pm}} \genredpar_s(u)\. \genL^s,
\]
where $\genredpar_s(u)$ are arbitrary polynomials in $u$. We also assume that 
$\genredpar_0(u) \ne 0$.
As before, the reduced derivation is obtained from the requirement that its bracket 
with $\genRL$ closes on the reduced algebra. This fixes the form of the derivation 
to be 
\[ \label{eq:derivation_genred}
\genRDL = \genODL + \sum_{\pm} \gamma_\pm(u) \.\genL^{\pm},
\]
with
\begin{align} \label{eq:genRedD}
\gamma_\pm(u) 
&= 
\pm \frac{u}{\genredpar_0(u)} \brk[s]*{\genredpar_\pm'(u)
- \frac{1}{2} \frac{\gamma'(u)}{\gamma(u)}\genredpar_\pm(u) },
\\
\gamma(u) 
&= 
\genredpar_0(u)^2 - 4 \genredpar_+(u) \genredpar_-(u)  
\end{align}
Again we observe the possibility of new poles apart from $0, \infty$. In order to 
tackle these, we have to either enlarge our algebra to an $n$-point algebra, where 
$n$ is the total number of poles, or remove the additional poles by 
adding distributional terms to \eqref{eq:genRedD}. Since it is not clear how to 
extend the construction to the coalgebra in the former option, we proceed with the 
latter approach, though we will not write the distributional terms explicitly
\footnote{As long as the contour of integration around $\infty$ is taken to be 
sufficiently small, the distributional terms do not make any contributions to the 
resulting expressions}.

The choice of the angular momentum generator direction naturally singles out an 
ideal within the momentum subalgebra: the orthogonal plane to $\genRL$ is rotated by 
the generator, while the parallel direction is unchanged. Thus, we can take a 
quotient of the momentum subalgebra by the orthogonal momentum directions.
In the presence of the affine extension, the ideal must be invariant under 
the shifted derivation \eqref{eq:derivation_genred}. Therefore, orthogonal 
momentum generators are dressed by the central charges. Namely, 
the ideal is spanned by the vectors 
\[
\gen{I}^\pm 
= 
\genredpar_0(u) \genP^\pm + 2 \genredpar_{\mp}(u) \genP^0 + \eta^\pm_u \genRCP,
\]
where the operator $\eta^\pm_u$ is defined to evaluate a function of $u$ 
that is multiplied to it to some number. 
In other words, we define $\eta^\pm_u u^n:= \eta^\pm_n$ 
as a shortcut notation to denote the coefficient $\eta^\pm_n$ applicable to the loop level $n$. 
Concretely, we set
\[
\eta^\pm_u f(u)
=
\mp\frac{1}{\pi\iunit}\oint_\infty f(u) 
\brk[s]*{\genredpar_\mp'(u)
- \frac{1}{2} \frac{\gamma'(u)}{\gamma(u)} \genredpar_\mp(u) } \diff{u}.
\]
Dividing out this ideal imposes the equivalence relations
\[
\genP^0 \simeq \genredpar_0(u) \genRP,
\eqsep
\genP^\pm \simeq - 2 \genredpar_\mp(u) \genRP 
- \eta^\pm_u \genredpar_0(u)^{-1}\genRCP.
\]

Altogether, the reduced algebra relations read
\begin{align}
\liebr!{g(u) \genRDL}{f(u) \genRL} 
&= 
u g(u) \brk[s]*{f'(u) 
+ \half f(u) \frac{\gamma'(u)}{\gamma(u)}}
\genRL
\\
&\alignrel
+ \frac{1}{\pi \iunit} \oint_\infty u g(u)
\frac{\genredpar_+'(u) \genredpar_-(u) - 
\genredpar_+(u) \genredpar_-'(u) }{\genredpar_0(u)} 
\\
&\qquad\qquad\cdot
\brk[s]*{f'(u) + \half f(u) \frac{\gamma'(u)}{\gamma(u)}}
\diff{u} \genRCL,
\\
\liebr!{g(u) \genRDP}{f(u) \genRL} 
&= 
u g(u)\gamma(u) \brk[s]*{f'(u) + \half f(u) \frac{\gamma'(u)}{\gamma(u)}}
\genRP
\\
&\alignrel
- \frac{1}{\pi \iunit} \oint_\infty u g(u)
\frac{ \genredpar_+'(u) \genredpar_-(u) 
- \genredpar_+(u) \genredpar_-'(u) }{\genredpar_0(u)} 
\\
&\qquad\qquad\cdot
\brk[s]*{f'(u) + \half f(u) \frac{\gamma'(u)}{\gamma(u)}}
\diff{u} \genRCP,
\\
\liebr!{g(u) \genRDP}{f(u) \genRL} 
&= 
u g(u) \brk[s]*{f'(u) - \half f(u) \frac{\gamma'(u)}{\gamma(u)}}
\genRP,
\\
\liebr!{f(u) \genRL}{g(u) \genRL} 
&= 
- \frac{1}{2 \pi \iunit} \oint_\infty f(u) \gamma(u)
\brk[s]*{g'(u)  + \half g(u)\frac{\gamma'(u)}{\gamma(u)}} \diff{u} \genRCL.
\\
\liebr!{f(u) \genRL}{g(u) \genRP} 
&= 
- \frac{1}{2 \pi \iunit} \oint_\infty f(u) \brk[s]*{g'(u) 
- \half g(u) \frac{\gamma'(u)}{\gamma(u)}} \diff{u} \genRCP.
\end{align}
%

\paragraph{Two-Level Reduction.} 

In the generic case it is very cumbersome to 
determine constraints on the r-matrix that are compatible with the reduction.

Here we give an explicit example of an r-matrix that can be obtained from the 
$\alg{iso}(2,1)$ r-matrix via a two-level reduction. The reduced $\genRL$ 
generator is given by \eqref{eq:genRedL} with the coefficients
\begin{align}
\genredpar_-(u) 
&=
\genredpar_{-,0} + \genredpar_{-,1} u + \genredpar_{-,2} u^2,
\\
\genredpar_0(u)
&=
\frac{\genredpar_{-,0} \genredpar_{0,2}}{\genredpar_{-,2}}
+ \frac{\genredpar_{-,1} \genredpar_{0,2}}{\genredpar_{-,2}} u
+ \genredpar_{0,2} u^2,
\\
\genredpar_+(u)
&=
\frac{\genredpar_{-,0} \genredpar_{+,1}}{\genredpar_{-,1}}
+ \genredpar_{+,1} u 
+ \frac{\genredpar_{-,2} \genredpar_{+,1}}{\genredpar_{-,1}} u^2.
\end{align}
This reduction produces the reduced r-matrix
\[
r(u,v) 
= 
\rnormL\frac{ \genredpar_-(v)}{ \genredpar_{-}(u)} \frac{1}{u-v} \genRL \otimes \genRP
+ \rnormL\frac{ \genredpar_-(u)}{ \genredpar_{-}(v)}\frac{1}{u-v} \genRP \otimes \genRL
+ \rnormP \ldots \genRP \otimes \genRP.
\]
Expanding the first two terms in $u/v$ around $0$ we obtain
\[
r(u,v) 
= 
\sum_{n \ge 0} \brk[s]*{\frac{u^n}{v^{n+1}} \rho_{n+2}(v) \genRL \otimes \genRP
+ \frac{1}{v^{n+3}} \tilde{\rho}_{n+2}(u) \genRP \otimes \genRL }
+ \rnormP \ldots \genRP \otimes \genRP,
\]
where $\rho_n, \tilde{\rho}_n$ are some polynomials of order $n$, which 
coefficients are expressed in terms of the $\genredpar$. 
This expansion induces the dualisation (we omit the terms $\sim \rnormP$)
\begin{align}
\brk!{\genRL_{n\ge0}}^* 
&\simeq 
\sum_{k=0}^{n+2} \rho_{n+2}^{(k)} \genRP_{k-n-1},
\\
\genRL_{n\le-3} 
&\simeq 
\brk*{\sum_{k=0}^{-n-1}\tilde{\rho}_{-n-1}^{(k)} \genRP_k}^*.
\end{align}
Now we observe, that $\genRP_{0,1}$ exist on both sides of the dualisation. 
Interestingly, $\genL_{-2,-1}$ do not appear in the r-matrix.

This illustrates that a consistent bialgebra reduction is not restricted to 
the level-1 reduction only. There are many possibilities to alter the choice of 
the reduced Lorentz generator that are compatible with the quasi-triangular 
bialgebra structure. It is interesting to understand what alternative reductions 
are admissible and to what models they should correspond.

\section{Supersymmetry}
\label{sec:rationalsusy}

In this section we discuss how our construction extends to the supersymmetric case 
relevant to the AdS/CFT integrability. Since the contraction 
and reduction mostly affect the bosonic part discussed above, lifting 
the procedure to the supersymmetric case is virtually straightforward.

\paragraph{Contraction.}

Now, our starting point is the affine Kac--Moody algebra $\affine{\alg{d}(2,1; \epsilon)}$ based on the simple 
superalgebra $\alg{d}(2,1; \epsilon)$.
\unskip\footnote{The conventional notation to denote this family of algebras is 
$\alg{d}(2,1; \redang)$. For subsequent construction we identify the parameter 
$\redang$ with the contraction variable $\epsilon$.}
We supplement this algebra with another $\affine{\alg{sl}(2)}$, such that the latter algebra together with an 
$\affine{\alg{sl}(2)}$ subalgebra within $\affine{\alg{d}(2,1; \epsilon)}$ play the role of the affine 
AdS algebra in the bosonic case. The contraction does not affect the supercharges 
and other bosonic generators in the exceptional superalgebra. Therefore we obtain 
an affine maximally extended $\affine{\alg{psu}(2|2)}_{\text{m.e.}}$ 
(with two sets of central charges and derivations).

We construct the rational r-matrix of the affine AdS superalgebra similarly to 
the bosonic case \eqref{eq:so22_r_matrix}: the Casimir operator $\casMM_1$ is 
now replaced by the invariant of $\alg{d}(2,1; \epsilon)$
\[
\casD = \casMM_1 + (\epsilon^{-1} - 1) \casJJ_\algleft - \epsilon^{-1} 
\casJJ_\algright - \epsilon^{-1} \casQQ,
\]
where we use $\casQQ$ to denote the anti-symmetric combination of the supercharges
\[
\casQQ
:= 
\half \varepsilon_{i j} \varepsilon_{l m} \varepsilon_{r p} 
\genQ^{i, l r} \otimes \genQ^{j, m p},
\]
with $\varepsilon$ being the anti-symmetric $2\times2$ matrix. The contraction of this r-matrix 
requires appropriate choice of all parameters as functions of $\epsilon$ 
\eqref{eq:contraction_coefficients}. This yields a finite r-matrix of the 
affine maximally extended $\alg{psu}(2|2)$ in the limit $\epsilon \to 0$.

\paragraph{Reduction.}

The reduction of the superalgebra $\affine{\alg{psu}(2|2)}_{\text{m.e.}}$ follows the 
same lines as for the bosonic subalgebra since the identification of 
the $\alg{u}(1)$ subalgebra and modding out the generated ideal does not affect 
the rest of the superalgebra. Let us only write down the non-trivial 
algebra relations that differ from those of $\affine{\alg{psu}(2|2)}_{\text{m.e.}}$ 
in the resulting deformed $\affine{\alg{u}(2|2)}$ algebra
\begin{align} \label{eq:susy_brackets}
\liebr!{g(u)\genRDL}{f(u) \genQ^{i, l r}} 
&= 
u g(u) f'(u) \genQ^{i, l r}+
u g(u) f(u) W_{\genD}(u)^i{}_j \genQ^{j, l r},
\\
\liebr!{g(u)\genRL}{f(u) \genQ^{i, l r}}
&=
g(u) f(u) W_{\genRL}(u)^i{}_j \genQ^{j, l r},
\\
\liebr!{g(u)\genQ^{i, l r}}{f(u) \genQ^{j, m p}}
&=
- 2 \mref^{-1} \brk!{W_{\genRL}(u) \varepsilon}{}^{i j} 
\varepsilon^{l m} \varepsilon^{r p} f(u) g(u) \genRP
\\
&\alignrel
- 2 \mref^{-1} \varepsilon^{l m} \varepsilon^{r p} 
\frac{1}{2 \pi \iunit} \oint_\infty 
\varepsilon^{i j} f(u)  \. \der g(u) \. \genRCP
\\
&\alignrel
- 2 \mref^{-1} \varepsilon^{l m} \varepsilon^{r p} 
\frac{1}{2 \pi \iunit} \oint_\infty 
\brk!{W_{\genD}(u) \varepsilon}{}^{i j} f(u)  g(u) \diff{u} \genRCP
\\
&\alignrel
+ \ldots,
\end{align}
where the ellipsis contains the canonical superalgebra terms in $\genJ_{\algleft,\algright}$. 
The matrices are defined as follows
\[
W_{\genD}(u) 
= \frac{1}{2} \frac{\redpar}{u^2 - \redpar^2}
\begin{pmatrix}
  0 & \iunit \eunit^{\iunit \redang} \\
  - \iunit \eunit^{-\iunit \redang} & 0
\end{pmatrix},
\eqsep
W_{\genRL}(u) 
= \frac{1}{2}
\begin{pmatrix}
  u \redpar^{-1} & \iunit \eunit^{\iunit \redang}\\
  \iunit \eunit^{-\iunit \redang}& -  u \redpar^{-1}
\end{pmatrix}.
\]

The reduction is also applied to the r-matrix of $\affine{\alg{psu}(2|2)}_{\text{m.e.}}$.
This again requires tuning of the parameter \eqref{eq:reduction_r_matrix_parameter_fix},
and one obtains the r-matrix of the deformed $\affine{\alg{u}(2|2)}$ given by 
\eqref{eq:rmat_reduced} enhanced by terms proportional to 
$\casJJ_{\algleft, \algright}$ and $\casQQ$. The phase of the r-matrix is determined 
by the reduction prescription up to one-parameter term $(u-v)\genRP\otimes \genRP$.

\paragraph{Classical Double.} 

The double construction can also be extended to the supersymmetric case 
without complications. The algebra $\alg{g}_+$ is now enlarged by the 
polynomials in the spectral parameter $u$ valued in 
$\Span\set{\genQ^{i, l r}, \genJ^a_{\algleft,\algright}}$. The dualisation is 
given by the standard relations mapping the polynomials to polynomials 
in $u^{-1}$ without a constant term. From the algebra relations 
\eqref{eq:susy_brackets} one can convince oneself that algebras $\alg{g}_+$ and 
$\brk{\alg{g}_+}^*$ close due to the fact that the reduction mixes generators 
of only one level difference in the affine algebra.

\paragraph{Affine Derivative.} 

It is interesting to compare our result to (the classical limit of) \cite{Borsato:2017lpf}.
In \cite{Borsato:2017lpf} the authors obtain a generator that behaves similarly to 
a derivation as the Lorentz boost of a q-deformed 1+1 dimensional Poincaré 
algebra. In our case, the supersymmetric cobracket of the derivation reads
\begin{align}
\cobra\brk!{\genRDL}
&=
-\rnormL\frac{u_1u_2+\redpar^2}{(u_1^2-\redpar^2)(u_2^2-\redpar^2)}
\brk!{\genRL\otimes\genRP-\genRP\otimes\genRL}
\\
&\alignrel
- \rnormL \mref \redpar \frac{\iunit \eunit^{- j \iunit \redang}
\brk!{(u_1^2-\redpar^2) - (u_2^2-\redpar^2)}}
{4 (u_1^2-\redpar^2) (u_1^2-\redpar^2) (u_1 - u_2)}
\delta_{ij} \varepsilon_{lm} \varepsilon_{rp} \genQ^{i,lr}\otimes\genQ^{j,mp}
\\
&\alignrel
-\rnormP\frac{(u_1^2-u_2^2)}{(u_1^2-\redpar^2)(u_2^2-\redpar^2)} \genRP \otimes \genRP,
\end{align}
which clearly differs from the cobracket of the Lorentz boost generator of 
\cite{Borsato:2017lpf}. In order to recover the connection to \cite{Borsato:2017lpf}, 
we discard the central charges and extend the definition of the cobracket to the 
Witt algebra of derivations 
(or at least to its $\alg{sl}(2)$ subalgebra) by the adjoint action of the derivations 
on the same r-matrix. The resulting expression 
\begin{align}
\cobra\brk!{\genRDL[m]} 
&= 
\rnormL \brk[s]*{
\frac{u_1^{m+2}}{\brk{u_1-u_2}\brk{u_1^2 - \redpar^2}}
-\frac{u_2^{m+2}}{\brk{u_1-u_2}\brk{u_2^2 - \redpar^2}}
}
\brk!{\genRL\otimes\genRP-\genRP\otimes\genRL} 
\\
&\alignrel
- \rnormL \frac{u_1^{m+1} - u_2^{m+1}}{(u_1 - u_2)^2} 
\brk!{\genRL\otimes\genRP+\genRP\otimes\genRL} 
\\
&\alignrel
+ \rnormL \mref \frac{u_1^{m+1} - u_2^{m+1}}{(u_1 - u_2)^2} 
\brk!{\casJJ_\algleft - \casJJ_\algright - \casQQ}
\\
&\alignrel
- \rnormL \frac{\iunit \eunit^{- j \iunit \redang} \mref \redpar }{4} \frac{
\brk!{u_2^{m+1}(u_1^2-\redpar^2) - u_1^{m+1} (u_2^2-\redpar^2)}}
{(u_1^2-\redpar^2) (u_1^2-\redpar^2) (u_1 - u_2)}
\delta_{ij} \varepsilon_{lm} \varepsilon_{rp} \genQ^{i,lr}\otimes\genQ^{j,mp}
\\
&\alignrel
+ \rnormP 
\brk[s]*{
\frac{u_1^{m+1}}{u_1^2-\redpar^2} 
-\frac{u_2^{m+1}}{u_2^2-\redpar^2}
-\frac{\brk{u_1 u_2 - \redpar^2}\brk{u_1^{m+1}-u_2^{m+1}}}
{\redpar^2 \brk{u_1-u_2}^2}    
}\genRP \otimes \genRP.
\end{align}
is superficially well-defined, since it is anti-symmetric and originates from 
the r-matrix satisfying the CYBE. However, for $m \ne -1$ the anti-symmetry only 
holds up to 
distributional terms that we omitted in the expression above. Equivalently, 
the symmetric part of the classical r-matrix is $\ad$-invariant only up to 
the aforementioned distributions. Nevertheless, we formally proceed with this 
bialgebra structure. Then we calculate the cobracket of the following combination 
of derivations $\genD_{+1} - \redpar^2 \genD_{-1}$
\begin{align}
\cobra\brk!{\genRDL[+1] - \redpar^2 \genRDL[-1]}
&=
-\rnormL\frac{2}{u_1-u_2}
\brk!{u_2 \genRL\otimes\genRP+u_1\genRP\otimes\genRL}
\\
&\alignrel
+\rnormL \mref \frac{u_1+u_2}{u_1 - u_2}
\brk!{\casJJ_\algleft - \casJJ_\algright - \casQQ}
\\
&\alignrel
-\rnormP\frac{u_1 u_2 - \redpar^2}{\redpar^2}\frac{u_1+u_2}{u_1-u_2} \genRP \otimes \genRP,
\end{align}
which perfectly matches the result obtained in \cite{Borsato:2017lpf}. Therefore 
at the classical level we can view the Lorentz boost generator as a particular 
element of the Witt algebra that acts on the bialgebra. However, the introduction 
of the affine charge singles out the particular derivation at level $-1$ for the 
bialgebra to be consistent (including distributional terms or in the form of loop levels). 
It is interesting to understand the role (if any) of the central extension in the AdS/CFT integrability.

\section{Trigonometric Case}
\label{sec:trigonometricbosonic}

In the previous sections we discussed the Lie bialgebra relevant to the AdS/CFT and 
the Hubbard model. This bialgebra has the classical r-matrix of the rational type. 
However, as it was mentioned in \secref{sec:gen_reduction}, there exist other 
possibilities for the reduction that also allow for other types of solutions 
for the CYBE. In this section we repeat the construction in the case of the 
trigonometric r-matrix, which is relevant to q-deformations of the AdS/CFT as well 
as of the Hubbard model. We perform the derivation of the affine bialgebra, 
discuss the evaluation representation and comment on the classical double construction.

\subsection{Trigonometric Reduction}

At first, we consider the reduction of the affine algebra. In order to simplify 
the expressions, we consider the parametric form of corresponding loop algebras with the 
loop parameter $z$. We also omit 
all the distributional terms, which remove all superficial poles at the points 
other than $z = 0, \infty$.

\paragraph{Reduced Derivation.}

For the trigonometric case we use the derivation of level-0 and throughout 
this section we denote $\genD := \genD_0$. As in the rational case, 
the action of the derivation on the reduced generator
\[
\genL =  
\half \redtmod^{-1} z \genL^0 
- \half \redtmod^{-1} \genL^0 
+ \ihalf \eunit^{\iunit \redang} \genL^- 
+ \ihalf \eunit^{- \iunit \redang} z \genL^+
\]
does not close on the affine $\alg{gl}(1) \times \Complex$
\[ \label{eq:derivationTrigonometric}
\comm!{\genODL}{f(z) \genL} 
= 
z f'(z) \genL 
+ \half \redtmod^{-1} z f(z) \genL^0 
+ \half \iunit \eunit^{-\iunit \redang} z f(z) \genL^+.
\]
Therefore, we supplement the derivation with the following combination 
of angular momentum generators
\[ \label{eq:reduced_derivation_trig}
\genRDL
:=
\genODL 
- \frac{2 \redtmod^2 z \genL^0 
+ \iunit \eunit^{-\iunit \redang} \redtmod z \genL^+
- \iunit \eunit^{ \iunit \redang} \redtmod z \genL^-}
{(z - \trigp{+}) (z - \trigp{-})} 
\]
where we define the following combinations of the reduction parameter $\redtmod$:
\[
\redtmod' = \sqrt{1 - \redtmod^2}, 
\eqsep 
\trigp{\pm} = (\iunit \redtmod \pm \redtmod')^2,
\eqsep
(z - \trigp{+}) (z - \trigp{-})=(z-1)^2+4\redtmod^2z
.
\]

We observe that the additional terms do not belong to the loop algebra 
$\alg{iso}(2,1)[z, z^{-1}]$, since the functions have additional poles at $z =\trigp{\pm}$. 
Therefore, one should consider a bigger (namely, a 4-point 
affine algebra) or introduce formal distributions from the algebra of formal 
power series
$\alg{iso}(2,1)[[z, z^{-1}]]$ in order to remove the new poles. In what follows 
we consider the latter approach (but we refrain from explicit display of the 
distributions), although it would be interesting to investigate the former 
possibility.

As before, the other derivation remains unchanged
\[
\genRDP := \genODP.
\]
%

\paragraph{Reduced Centre.}

As in the rational case we add a central charge to the ideal generators of 
the trigonometric reduction
\begin{align} \label{eq:ideal_trigonometric_affine}
\gen{I}^0
&:=
\ihalf \eunit^{\iunit \redang} \genP^- 
- \ihalf \eunit^{-\iunit \redang} z \genP^+  
+ \eta^0_z \genRCP,
\\
\gen{I}^+
&:=
\iunit \eunit^{\iunit \redang} \genP^0 
+ \half \redtmod^{-1} (z - 1) \genP^+ 
+ \eta^+_z \genRCP.
\end{align}
Again, the operators $\eta^{0,+}_z$ are functionals on the space 
of Laurent polynomials in $z$ and we fix their action by requiring that $
\gen{I}^{0,+}$ span an ideal w.r.t.\ $\genRDL$ and $\genRL$
\begin{align}
\eta^0_z f(z)
&=
\frac{\redtmod}{2 \pi \iunit} 
\oint_\infty \frac{f(z) (z + 1)}{(z - \trigp{+}) (z - \trigp{-})} \diff{z},
\\
\eta^+_z f(z) 
&=
\frac{\iunit \eunit^{\iunit \redang}}{2 \pi \iunit}
\oint_\infty \frac{ f(z) (z + 2\redtmod^2-1) }
{(z - \trigp{+}) (z - \trigp{-})}  \diff{z}.
\end{align}
Modding out this ideal amounts to the identifications among momentum generators 
\begin{align} \label{eq:trig_red_rels}
\genP^0 
&= 
\half \redtmod^{-1} (z - 1) (\genRP - \eta^{\gaugeDLL}_z \genRCP)
+ \iunit \eunit^{-\iunit \redang} \eta^+_z \genRCP,
\\
\genP^+
&=
-\iunit \eunit^{\iunit \redang}  (\genRP - \eta^{\gaugeDLL}_z \genRCP),
\\
\genP^-
&=
-\iunit \eunit^{-\iunit\redang} z  (\genRP - \eta^{\gaugeDLL}_z \genRCP) 
+ 2 \iunit \eunit^{-\iunit \redang} \eta^0_z \genRCP,
\end{align}
where the functional coefficient $\eta^{\gaugeDLL}_z$ is fixed to be
\[ \label{eq:gauge_fix_CP}
\eta^{\gaugeDLL}_zf(z) = 
-\frac{\redtmod}{ \pi \iunit} 
\oint_\infty \frac{f(z)}{(z - \trigp{+}) (z - \trigp{-})} \diff{z},
\]
in order to simplify the resulting algebra relations.

\paragraph{Resulting Algebra.}

To simplify formulae we define the quantity
\[
\trigV(z)
=
- \frac{z (z + 2\redtmod^2 - 1)}{(z - \trigp{+})(z - \trigp{-})},
\]
which is a logarithmic derivative of the eigenvalue of $\genRP$:
\[
z \frac{\der \log\genRP}{\der z} = \trigV(z).
\]
Altogether we have the following derivations
\begin{align} \label{eq:parametricDLDPrelations_trig}
\comm!{\genRDL}{f(z) \genRL}
&=
\brk!{z f'(z) - \trigV(z) f(z)} \genRL,
\\
\comm!{\genRDL}{f(z) \genRP}
&=
\brk!{z f'(z) + \trigV(z) f(z) } \genRP,
\\
\comm!{\genRDP}{f(z) \genRL}
&=
\frac{(z - \trigp{+})(z - \trigp{-})}{4 \redtmod^2} 
\brk!{z f'(z) - \trigV(z) f(z)} \genRP.
\end{align}
The non-trivial commutators between the $\genRL$ and $\genRP$ read
\begin{align} \label{eq:parametricLLLPrelations_trig}
\comm!{f(z) \genRL}{ g(z) \genRL}
&=
- \frac{1}{2 \pi \iunit} 
\oint_{\infty} \brk!{z g'(z) - \trigV(z) g(z)} f(z) 
\frac{(z - \trigp{+})(z - \trigp{-})}{4 \redtmod^2 z} \diff{z} \genRCL,
\\
\comm!{f(z) \genRL}{ g(z) \genRP}
&=
- \frac{1}{2 \pi \iunit} \oint_{\infty} \brk!{z g'(z) + \trigV(z) g(z)} 
\frac{f(z)}{z} \diff{z}  \genRCP.
\end{align}
Notice that these relations essentially extend those obtained in 
\cite{Beisert:2007ty} by introducing the second set of the affine derivation and 
central charge.

\paragraph{Evaluation Representation.}

In the trigonometric case the 2-parameter evaluation representation space of 
the affine 3D Poincaré algebra is spanned by the states 
$\state{z, y, p, \phi}_{m,s}$ and the action of the generators is 
given by the differential operators
\begin{align} \label{eq:two_param_rep_trig}
\rho_{z, y}\brk!{f(z) \genL^a}
&= 
f(z) \rho(\genL^a)  + f'(z) z y \rho(\genP^a),
&
\rho_{z, y}\brk!{f(z) \genP^a} 
&= 
f(z) \rho(\genP^a),
\\
\rho_{z, y}\brk!{f(z) \genODL}
&= 
- f(z) z \frac{\partial}{\partial z}
- f'(z) z y \frac{\partial}{\partial y} ,
&
\rho_{z, y}\brk!{f(z) \genODP} 
&=
- \frac{\partial}{\partial y} ,
\end{align}
where $\rho$ stands for the representation \eqref{eq:momrep} 
of the level-0 algebra $\alg{iso}(2,1)$.

The reduction identifications \eqref{eq:trig_red_rels} can be conveniently 
resolved for the eigenvalues of $\genP^a$ in the following parametrisation
\[
e_m(p) = (z-1)\frac{qm}{\redtmod},
\eqsep
p \mathinner{\eunit}^{\iunit \phi} = -2 \iunit \mathinner{\eunit}^{\iunit\redang}qm,
\eqsep
p \mathinner{\eunit}^{- \iunit \phi} = -2\iunit \mathinner{\eunit}^{-\iunit\redang}qzm.
\]
The parameters $q$ and $z$ can be expressed in terms of a uniform variable $x$
\[
z=\frac{\iunit x}{(\redtmod'x-\iunit \redtmod)(\redtmod x+\iunit \redtmod')},
\eqsep
q=\frac{(\redtmod'x-\iunit \redtmod)(\redtmod x+\iunit \redtmod')}{\redtmod'(x^2-1)},
\]
such that the mass shell constraint for the vector $\genP$ is explicitly satisfied.
Therefore, it is natural to view the variable $x$ as one of the two evaluation 
representation parameters of the reduced representation space instead of $z$:
\[
    \state{x,y}_{m,s} := \zeta(x) \state!{z(x), y, p(x), \phi(x)}_{m,s},
\]
where we also rescale the state by a potentially non-trivial function of $x$.

For the reduced algebra we would like that the representation of the derivations 
$\genRDL$ and $\genRDP$ acts as total derivatives w.r.t.\ the parameters $z$ and 
$y$, i.e.
\[
\rho_{z, y}(\genRDL) \state{x,y}_{m,s}
\overset{!}{=}
- z \frac{\partial x}{\partial z}\frac{\der}{\der x} \state{x,y}_{m,s},
\eqsep
\rho_{z, y}(\genRDP) \state{x,y}_{m,s}
\overset{!}{=}
- \frac{\der}{\der y} \state{x,y}_{m,s}.
\]
The latter is trivially satisfied. For the former, 
we explicitly evaluate the l.h.s.
\begin{align}
\rho_{z, y}(\genRDL) \state{x,y}_{m,s}
&=  
\zeta(x) \brk*{
-  \ihalf \frac{\partial}{\partial \phi} 
+  \iunit m  q(z) \sqrt{z} \brk!{1 + 2 \trigV(z)}
\frac{\partial}{\partial p}}\state{z, y, p, \phi}_{m,s}
\\
&\alignrel
+\zeta(x) \brk*{- z \frac{\partial }{\partial z} 
+ \zeta(x) \frac{s \brk{q (z - 1) - \redtmod} }{2\redtmod} }
\state{z, y, p, \phi}_{m,s}.
\end{align}
We notice that the following relations hold
\[
z \frac{\partial \phi(z)}{\partial z}
=
\frac{\iunit}{2},
\eqsep 
z \frac{\partial p(z)}{\partial z}
=
 - \iunit m  q(z) \sqrt{z} \brk!{1 + 2 \trigV(z)},
\]
which allows us to write the representation of the derivation as
\begin{align}
\rho_{z, y}(\genRDL) \state{x,y}_{m,s}
&= 
 \zeta(x) \brk[s]*{- z \frac{\der}{\der z} 
 + \frac{s}{2\redtmod} \brk!{q (z - 1) - \redtmod}}  \state{z, y, p, \phi}_{m,s}.
\end{align}
Requiring that the r.h.s.\ is a total derivative w.r.t.\ $z$ gives an equation on 
$\zeta(x)$ that is resolved by 
\[
\zeta(x) = \brk{\redtmod x + \iunit \redtmod'}^{-s}.
\]
%

\paragraph{Supersymmetry.}

The trigonometric reduction can be also performed for the Poincaré supersymmetry.
Compared to the rational case, the action of the reduced generators
$\genRL$ and $\genRDL$ on the supercharges is given by the same relations 
\eqref{eq:susy_brackets}, 
but with the matrices $W_{\genRL}$ and $W_{\genD}$ replaced by 
\[
W_{\genRL}(z) 
=\frac{1}{4 \redtmod}
\begin{pmatrix}
  z-1 & 2 \redtmod \eunit^{\iunit \redang} \\
  2 \redtmod z \eunit^{-\iunit \redang} & 1 - z
\end{pmatrix},
\eqsep
W_{\genD}(z) 
=
\frac{\redtmod}{(z - \trigp{+})(z - \trigp{-})}
\begin{pmatrix}
  - \redtmod & 
  \eunit^{\iunit \redang}  \\
  -\eunit^{-\iunit \redang} & 
  \redtmod
\end{pmatrix}.
\]
%

\subsection{r-Matrix and Coalgebra}

Now we are in the position to perform the reduction of the coalgebra structure.
After applying the contraction and reduction procedures on 
the standard r-matrix of the affine 
$\affine{\alg{sl}(2)} \times \affine{\alg{sl}(2)}$
we obtain the r-matrix 
\begin{align} \label{eq:rmat_u1R_trig}
r_{\text{trig}} 
&= 
\rnormL \frac{z_2}{z_1 - z_2} \genL \otimes \genP + 
\rnormL \frac{z_1 }{z_1 - z_2} \genP \otimes \genL
\\
&\alignrel
+  \rnormP \frac{\rfrac{1}{8} \redtmod^{-2} (z_1 + z_2) (z_1 - 1) (z_2 - 1) + 
z_1 z_2}{z_1 - z_2} \genP \otimes \genP
\\
&\alignrel
+ \rnormL \genRCL \otimes \genRDP + \rnormL \genRCP \otimes \genRDL 
+ \rnormP \genRCP \otimes \genRDP.
\end{align}
The r-matrix induces a coalgebra structure for the
resulting deformed affine $\alg{u}(1) \times \alg{\Real}$:
\begin{align}
\cobra(\genRDL)
&=
- \rnormL \frac{(z_1 - 1)(z_2 - 1) - 2 \redtmod^2 (1 + z_1 z_2)}
{(z_1 - \trigp{+})(z_1 - \trigp{-})(z_2 - \trigp{+})(z_2 - \trigp{-})} 
\brk*{z_2 \genRL \otimes \genRP - z_1 \genRP \otimes \genRL}
\\
&\alignrel
+ \rnormP \frac{(z_1 + 1)(z_2 + 1) ( z_1 - z_2)(z_1 z_2 - 1)}
{4 (z_1 - \trigp{+})(z_1 - \trigp{-})(z_2 - \trigp{+})(z_2 - \trigp{-})}
\genRP \otimes \genRP,
\\
\cobra(\genRDP)
&= 
0
\\
\cobra\brk!{f(z) \genRL}
&=
- \rnormL \genRCP \wedge 
\bigg[ 
z f_+'(z) - \trigV(z) f_+(z) 
- \frac{\trigp{+}}{2}
\frac{f_+(\trigp{+})}{z - \trigp{+}} 
-\frac{\trigp{-}}{2}
\frac{f_+(\trigp{-})}{z - \trigp{-}}
\bigg] \genRL
\\
&\alignrel
- \rnormL \genRCL \wedge
\bigg[ 
\frac{(z - \trigp{+})(z - \trigp{-})}{4 \redtmod^2} \brk!{z f_+'(z) - \trigV(z) f_+(z)}
\bigg] \genRP
\\
&\alignrel
- \rnormP \genRCP \wedge 
\bigg[
\frac{(z - \trigp{+})(z - \trigp{-})}{4 \redtmod^2} \brk!{z f_+'(z) - \trigV(z) f_+(z)}
\\
&\qquad\quad
- \frac{\trigp{\pm}}{16\redtmod^2} \frac{1}{z - \trigp{\pm}}
\big[(z-1)\brk!{\trigp{\pm}f_+(\trigp{\pm}) + f_{-1}} 
\\
&\qquad\qquad
+ \brk!{(z-1)^2 + 8 \redtmod^2 z}f_+(\trigp{\pm})
+ z(z-1)(\trigp{\pm})^{-1}\brk!{f_+(\trigp{\pm}) -  f_0} \big]
\\
&\qquad\quad
- \frac{z-1}{8 \redtmod^2} \brk!{f_0 \trigV(z) + f_{-1} (\trigV(z)+1)}
\bigg] \genRP,
\\
\cobra\brk!{f(z) \genRP}
&=
- \rnormL \genRCP \wedge 
\bigg[
z f_+'(z) + \trigV(z) f_+(z) 
+ \frac{z}{2}\frac{f_+(\trigp{+})}{z - \trigp{+}}
+ \frac{z}{2}\frac{f_+(\trigp{-})}{z - \trigp{-}}
\bigg] \genRP,
\end{align}
where as before $f_+(z)$ denotes a projection of the Laurent polynomial 
$f(z) \in \Complex[z, z^{-1}]$
on the non-negative modes in $\Complex[z]$ and $f_n$ denotes its $n$-th coefficient.

\paragraph{Rational Limit.}

We can verify the formulae above by considering the rational limit 
$\ratlimpar \to 0$ 
\[
z_i = \eunit^{\ratlimpar u_i}, 
\eqsep 
y_i = \ratlimpar v_i,
\eqsep 
\redtmod = \ihalf \ratlimpar \redpar.
\]
The loop part of the r-matrix goes to that of the rational r-matrix 
(cf. \cite{Beisert:2022vnc}). Using the relations 
\[
z_i - \trigp{\pm} = (u_i \pm \redpar) \ratlimpar  + \Order(\ratlimpar^2),
\eqsep 
z_i + 2\redtmod^2 - 1 = u_i \ratlimpar  + \Order(\ratlimpar^2),
\]
we also verify that the limit is consistent for the derivation:
\[ \label{eq:rat_limit_derivation}
\genRDL 
=
\frac{1}{\ratlimpar u} \brk*{ - u \frac{\partial}{\partial u} 
+ \frac{\redpar u}{u^2 - \redpar^2} 
\brk!{\half \eunit^{- \iunit \redang} \genL^+ 
- \half\eunit^{ \iunit \redang } \genL^-}}
+ \Order(\ratlimpar^0),
\]
where on the r.h.s.\ we recover precisely the derivation of the 
rational reduction (albeit scaled by $\ratlimpar^{-1}$).
One can easily see that \eqref{eq:rat_limit_derivation} induces a consistent reduction of the r-matrix
scaled by $\ratlimpar^{-1}$. Moreover, the algebra relations 
\eqref{eq:parametricDLDPrelations,eq:parametricLLLPrelations} can be obtained 
as a rational limit of 
\eqref{ eq:parametricDLDPrelations_trig, eq:parametricLLLPrelations_trig}.

\paragraph{Classical Double.}

Similarly to the rational case, we can obtain the algebras of interest as a 
classical double, albeit with minor modifications \cite{Chari:1994pz}. We 
consider the polynomial algebra with a central charge 
\[
\alg{g}_+ = 
\alg{sl}(2)[z]z \oplus \Complex \genJ^- \oplus \Complex \genJ^0 \oplus \Complex \genC.
\] 
The dual algebra $\alg{g}_- = \brk{\alg{g}_+}^*$ is induced by the dualisation
\[
\brk{\genJ^0}^* 
\simeq
- \half \rnorm \genJ^0,
\eqsep
\brk{\genJ^-}^* 
\simeq 
\half \rnorm \genJ^+,
\eqsep
\brk{z^{n>0}\genJ^a}^* 
\simeq
\rnorm c_{ab} z^{-n} \genJ^b,
\eqsep
\genC^* 
\simeq
\rnorm \genD.
\]
One can convince oneself that Lie brackets inherited from $\affine{\alg{sl}(2)}$
close on both algebras and produce a cobracket satisfying the 1-cocycle condition. 

Now, we are in the position to construct the classical double 
$\alg{g} = \alg{g}_+ \oplus \alg{g}_-$. However, the resulting algebra is slightly 
bigger than the affine $\affine{\alg{sl}(2)}$: the generator $\genJ^0$ appears 
on both sides of the dualisation. This can be cured by noticing that the combination
$ 2 \brk{\genJ^0}^* + \rnorm \genJ^0 $ is central and thus can be divided out
\[
\frac{\alg{g}}{\brk[a]{2 \brk{\genJ^0}^* + \rnorm \genJ^0}} = \affine{\alg{sl}(2)}.
\]
The classical r-matrix obtained from this dualisation is indeed the standard 
trigonometric r-matrix. 

Following the \secref{sec:rat_double}, we can apply the contraction and reduction 
to the trigonometric double construction. Omitting the intermediate steps, the 
final dualisation can be read off from the r-matrix \eqref{eq:rmat_u1R_trig}.
\unskip\footnote{Keep in mind that all functions with poles at points other than 
$0, \infty$ has to be expanded in formal power series around $z = \infty$.}
We 
notice that the classical double resulting from this dualisation is again slightly 
bigger than the required algebra: one has to divide out an ideal generated by 
$8 \redtmod^2(\genRP)^* + \rnormP (1 - z) \genRP$. 

The classical double construction also applies to the supersymmetric extension. 
In this case the ideal to be modded out is supplemented by the vectors 
$ 2 \brk{\genJ_{\algleft,\algright}^0}^* + \rnorm \genJ_{\algleft,\algright}^0$.
The supercharges split into $\alg{g}_+$ and $\alg{g}_-$ completely as they do 
not contribute to the Cartan subalgebra.

\section{Conclusions and Outlook}
\label{sec:conclusions}

In this paper we have constructed a classical affine Lie bialgebra for
AdS/CFT integrability and the one-dimensional Hubbard model by applying 
a contraction and reduction procedure (essentially, non-invertible Lie bialgebra homomorphisms) 
to $\affine{\alg{sl}(2)} \times \affine{\alg{d}(2,1;\alpha)}$. The resulting affine 
bialgebra has a peculiar feature of carrying two loop parameters, one of which 
can be viewed as infinitesimal. Correspondingly, there are two sets of affine 
derivations and central charges
(which may or may not serve a yet-to-be-understood purpose within the AdS/CFT context). 
The coalgebra also carries a non-standard feature
w.r.t.\ the affine structure: the cobracket of one of the 
derivations is non-zero, which can be attributed to the fact that the obtained 
classical r-matrices are of non-difference form. 
The analysis is performed for both rational and trigonometric solutions to the 
CYBE (the latter is relevant for q-deformed AdS/CFT \cite{Beisert:2008tw, Beisert:2010kk, Beisert:2011wq} 
and the Hubbard model \cite{Delduc:2013qra, Delduc:2014kha}). 
We also demonstrated that the resulting bialgebras can be 
implemented as classical doubles.

The extension of the reduction to the affine structure leads to the natural 
appearance of two additional poles (apart from the usual ones at $0$ and $\infty$) in 
the algebra. This hints towards 4-point Lie algebras \cite{Bremner:1994}. 
However, the Lie bialgebra structure in this case is unknown. It is interesting 
to understand if a classical r-matrix can be constructed in this case and 
what possible implications the additional poles might have for the AdS/CFT 
integrability.

Also, the notion of the reduction seems to have a lot of freedom. By an explicit 
example we demonstrated that other choices of ideals in the reduction 
that are compatible with the quasi-triangular bialgebra structure are possible. 
Therefore it is important to understand the amount of freedom we have to deform 
the reduction relations and what classical r-matrices one might obtain.

Curiously, the affine derivation seems to implement the q-deformed Poincaré boost 
generator \cite{Young:2007wd} at the classical level \cite{Borsato:2017lpf}. 
We argue that the exact form of the Lorentz boost symmetry is given by a particular 
element of a Witt algebra that acts on the reduced bialgebra. Interestingly, 
the affine extension naturally selects a slightly different element of the 
Witt algebra. Nevertheless, this derivation shows that a possible origin of the 
Lorentz boost symmetry is a quantum affine algebra, which is along the lines of 
\cite{deLeeuw:2011fr}, where the secret symmetry is also naturally derived from 
the quantum affine algebra of the q-deformed Hubbard model. 

All this naturally leads to a possible implementation of the quantum symmetry 
of the AdS/CFT S-matrix as a (rational limit of) quantum affine algebra. Following 
the logic of the classical analysis presented in this paper, the 
possible starting point in the quantum setting would be to consider the 
universal R-matrix of the quantum affine $\envalg_q(\affine{\alg{sl}(2)})$
\cite{Tolstoy:1992}. Then one has to lift the notion of the contraction and 
reduction to the quantum level. For the q-deformed non-affine $\envalg_q(\alg{sl}(2))$ and 
$\envalg_q(\alg{d}(2,1;\alpha))$ the contraction has already been implemented 
in \cite{Beisert:2016qei} leading to a kappa-deformed Poincaré algebra. Therefore, the next 
natural step is to generalise it to the affine algebras and develop the notion 
of the reduction in the quantum case. Of course, for the full symmetry algebra 
one would also have to consider the quantum affine exceptional superalgebra 
$\envalg_q(\affine{\alg{d}(2,1;\alpha)})$ \cite{Heckenberger:2007ry}, for which 
the universal R-matrix is unknown. Finding the quasi-triangular structure 
is then also crucial for the final construction.

\pdfbookmark[1]{Acknowledgements}{ack}
\subsection*{Acknowledgements}

We thank Riccardo Borsato and Alessandro Torrielli for discussions related to the work 
and reading the draft of this manuscript. EI thanks Raschid Abedin for interesting 
and useful discussions.
The work of NB and EI is supported by the Swiss National Science Foundation 
through the NCCR SwissMAP.

\ifarxiv\else
\begin{bibtex}[\jobname]

  @article{Beisert:2007ty,
  author = "Beisert, Niklas and Spill, Fabian",
  title = "The Classical r-matrix of AdS/CFT and its Lie Bialgebra Structure",
  eprint = "0708.1762",
  archivePrefix = "arXiv",
  primaryClass = "hep-th",
  reportNumber = "AEI-2007-116, HU-EP-07-31",
  doi = "10.1007/s00220-008-0578-2",
  journal = "Commun. Math. Phys.",
  volume = "285",
  pages = "537--565",
  year = "2009"
}

@article{Beisert:2010kk,
  author = "Beisert, Niklas",
  title = "The Classical Trigonometric r-Matrix for the Quantum-Deformed Hubbard Chain",
  eprint = "1002.1097",
  archivePrefix = "arXiv",
  primaryClass = "math-ph",
  reportNumber = "AEI-2010-016",
  doi = "10.1088/1751-8113/44/26/265202",
  journal = "J. Phys. A",
  volume = "44",
  pages = "265202",
  year = "2011"
}

@article{Borsato:2017lpf,
  author = "Borsato, Riccardo and Torrielli, Alessandro",
  title = "{$q$-Poincaré supersymmetry in AdS$_5$ / CFT$_4$}",
  eprint = "1706.10265",
  archivePrefix = "arXiv",
  primaryClass = "hep-th",
  reportNumber = "DMUS-MP-17-07, NORDITA-2017-066",
  doi = "10.1016/j.nuclphysb.2018.01.017",
  journal = "Nucl. Phys. B",
  volume = "928",
  pages = "321--355",
  year = "2018"
}

@article{Bremner:1994, 
  title={Generalized Affine Kac-Moody Lie Algebras Over Localizations of the Polynomial 
  Ring in One Variable}, 
  volume={37}, 
  DOI={10.4153/CMB-1994-004-8}, 
  number={1}, 
  journal={Can. Math. Bull.}, 
  publisher={Cambridge University Press}, 
  author={Bremner, Murray}, 
  year={1994}, 
  pages={21–28}
}

@article{Bremner:1995,
  ISSN = {00029939, 10886826},
  doi = {10.2307/2160931},
  author = {Murray Bremner},
  journal = {Proc. Am. Math. Soc.},
  number = {7},
  pages = {1981--1989},
  publisher = {American Mathematical Society},
  title = {Four-Point Affine Lie Algebras},
  urldate = {2023-01-25},
  volume = {123},
  year = {1995}
}

@article{Hartwig:2007,
  title = {The Tetrahedron algebra, the Onsager algebra, and the sl$_2$ loop algebra},
  journal = {J. Alg.},
  volume = {308},
  number = {2},
  pages = {840-863},
  year = {2007},
  issn = {0021-8693},
  doi = {https://doi.org/10.1016/j.jalgebra.2006.09.011},
  author = {Brian Hartwig and Paul Terwilliger},
  eprint = "0511004",
  archivePrefix = "arXiv",
  primaryClass = "math-ph",
}

@article{Beisert:2022vnc,
    author = "Beisert, Niklas and Im, Egor",
    title = "{Classical Lie Bialgebras for AdS/CFT Integrability by Contraction and Reduction}",
    eprint = "2210.11150",
    archivePrefix = "arXiv",
    primaryClass = "hep-th",
    doi = "10.21468/SciPostPhys.14.6.157",
    journal = "SciPost Phys.",
    volume = "14",
    pages = "157",
    year = "2023"
}

@article{Stolin:1990,
 ISSN = {00255521, 19031807},
 URL = {http://www.jstor.org/stable/24492600},
 author = {A. Stolin},
 journal = {Math. Scand.},
 number = {1},
 pages = {57--80},
 publisher = {Mathematica Scandinavica},
 title = {On Rational Solutions of Yang-Baxter Equation for $\alg{sl}(n)$},
 urldate = {2023-05-12},
 volume = {69},
 year = {1991}
}

@Book{Chari:1994pz,
     author    = "Chari, V. and Pressley, A.",
     title     = "A guide to quantum groups",
     address   = "Cambridge, UK",
     year      = "1994",
     pages     = "651",
     publisher = "Cambridge University Press"
}

@book{Klimyk:1997eb,
    author = "Klimyk, A. and Schmudgen, K.",
    title = "{Quantum groups and their representations}",
    year = "1997",
    doi = "10.1007/978-3-642-60896-4",
    pages     = "552",
    publisher = "Springer Berlin, Heidelberg"
}

@article{Young:2007wd,
    author = "Young, C. A. S.",
    title = "{q-deformed supersymmetry and dynamic magnon representations}",
    eprint = "0704.2069",
    archivePrefix = "arXiv",
    primaryClass = "hep-th",
    reportNumber = "DCPT-07-09",
    doi = "10.1088/1751-8113/40/30/033",
    journal = "J. Phys. A",
    volume = "40",
    pages = "9165--9176",
    year = "2007"
}

@article{Belavin:1982,
    author = "Drinfel'd, Vladimir Gershonovich and Belavin, Aleksandr Abramovich",
    title = "Solutions of the classical Yang-Baxter equation for simple Lie algebras",
    journal = "Func. Anal. Appl.",
    volume = "16",
    issue = "3",
    pages = "159--180",
    year = "1982",
    doi = "10.1007/BF01081585"
}

@misc{Abedin:2021,
      title={Geometrization of solutions of the generalized classical Yang-Baxter 
        equation and a new proof of the Belavin-Drinfeld trichotomy}, 
      author={Raschid Abedin},
      year={2022},
      eprint={2107.10722},
      archivePrefix={arXiv},
      primaryClass={math.AG}
}

@article{Drinfel'd:1985,
    author = "Drinfel'd, Vladimir Gershonovich",
    title = "Hopf algebras and the quantum Yang–Baxter equation",
    journal   = "Sov. Math. Dokl.",
    volume    = "32",
    pages     = "254-258",
    year      = "1985"
}

@article{Drinfel'd:1988,
    author = "Drinfel'd, Vladimir Gershonovich",
    title = "Quantum groups",
    doi = "10.1007/BF01247086",
    journal = "J. Sov. Math.",
    volume = "41",
    issue = "2",
    pages = "898--915",
    year = "1988"
}

@article{Hubbard:1963,
    ISSN = {00804630},
    URL = {http://www.jstor.org/stable/2414761},
    author = {J. Hubbard},
    journal = {Proc. R. Soc. London A},
    number = {1365},
    pages = {238--257},
    publisher = {The Royal Society},
    title = {Electron Correlations in Narrow Energy Bands},
    urldate = {2022-09-21},
    volume = {276},
    year = {1963}
}

@Book{Essler:2005aa,
     author    = "Essler, F. H. L. and Frahm, H. and Göhmann, F. and Klümper, A. and Korepin, V. E.",
     title     = "The one-dimensional Hubbard model",
     address   = "Cambridge, UK",
     year      = "2005",
     pages     = "690",
     publisher = "Cambridge University Press"
}

@Article{Shastry:1986bb,
    author = "Shastry, B. Sriram",
    title     = "Exact Integrability of the One-Dimensional Hubbard Model",
    journal   = "Phys. Rev. Lett.",
    volume    = "56",
    year      = "1986",
    pages     = "2453-2455",
    doi = "10.1103/PhysRevLett.56.2453",
}

@article{Beisert:2010jr,
    author = "Beisert, Niklas and others",
    title = "{Review of AdS/CFT Integrability: An Overview}",
    eprint = "1012.3982",
    archivePrefix = "arXiv",
    primaryClass = "hep-th",
    reportNumber = "AEI-2010-175, CERN-PH-TH-2010-306, HU-EP-10-87, HU-MATH-2010-22, KCL-MTH-10-10, UMTG-270, UUITP-41-10",
    doi = "10.1007/s11005-011-0529-2",
    journal = "Lett. Math. Phys.",
    volume = "99",
    pages = "3--32",
    year = "2012"
}

@article{Bombardelli:2016rwb,
    author = "Bombardelli, Diego and others",
    title = "{An integrability primer for the gauge-gravity correspondence: An introduction}",
    eprint = "1606.02945",
    archivePrefix = "arXiv",
    primaryClass = "hep-th",
    reportNumber = "CNRS-16-03, DCPT-16-19, DESY-16-083, DMUS-MP-16-09, HU-EP-16-13, NORDITA-2016-33, HU-MATH-16-08",
    doi = "10.1088/1751-8113/49/32/320301",
    journal = "J. Phys. A",
    volume = "49",
    number = "32",
    pages = "320301",
    year = "2016"
}

@article{Maldacena:1997re,
    author = "Maldacena, Juan Martin",
    title = "{The Large N limit of superconformal field theories and supergravity}",
    eprint = "hep-th/9711200",
    archivePrefix = "arXiv",
    reportNumber = "HUTP-97-A097, HUTP-98-A097",
    doi = "10.4310/ATMP.1998.v2.n2.a1",
    journal = "Adv. Theor. Math. Phys.",
    volume = "2",
    pages = "231--252",
    year = "1998"
}

@article{Minahan:2002ve,
    author = "Minahan, J. A. and Zarembo, K.",
    title = "{The Bethe ansatz for N = 4 superYang-Mills}",
    eprint = "hep-th/0212208",
    archivePrefix = "arXiv",
    reportNumber = "UUITP-17-02, ITEP-TH-73-02",
    doi = "10.1088/1126-6708/2003/03/013",
    journal = "JHEP",
    volume = "03",
    pages = "013",
    year = "2003"
}

@article{Beisert:2004ry,
    author = "Beisert, Niklas",
    title = "{The Dilatation operator of N = 4 super Yang-Mills theory and integrability}",
    eprint = "hep-th/0407277",
    archivePrefix = "arXiv",
    reportNumber = "AEI-2004-057",
    doi = "10.1016/j.physrep.2004.09.007",
    journal = "Phys. Rept.",
    volume = "405",
    pages = "1--202",
    year = "2004"
}

@article{Beisert:2003ys,
    author = "Beisert, Niklas",
    title = "{The su(2$/$3) dynamic spin chain}",
    eprint = "hep-th/0310252",
    archivePrefix = "arXiv",
    reportNumber = "AEI-2003-087",
    doi = "10.1016/j.nuclphysb.2003.12.032",
    journal = "Nucl. Phys. B",
    volume = "682",
    pages = "487--520",
    year = "2004"
}

@article{Beisert:2005fw,
    author = "Beisert, Niklas and Staudacher, Matthias",
    title = "{Long-range psu(2,2$/$4) Bethe Ansatze for gauge theory and strings}",
    eprint = "hep-th/0504190",
    archivePrefix = "arXiv",
    reportNumber = "AEI-2005-092, PUTP-2159",
    doi = "10.1016/j.nuclphysb.2005.06.038",
    journal = "Nucl. Phys. B",
    volume = "727",
    pages = "1--62",
    year = "2005"
}

@article{Beisert:2006qh,
    author = "Beisert, Niklas",
    title = "{The Analytic Bethe Ansatz for a Chain with Centrally Extended su(2$/$2) Symmetry}",
    eprint = "nlin/0610017",
    archivePrefix = "arXiv",
    reportNumber = "AEI-2006-074, PUTP-2211",
    doi = "10.1088/1742-5468/2007/01/P01017",
    journal = "J. Stat. Mech.",
    volume = "0701",
    pages = "P01017",
    year = "2007"
}

@article{Arutyunov:2007tc,
    author = "Arutyunov, Gleb and Frolov, Sergey",
    title = "{On String S-matrix, Bound States and TBA}",
    eprint = "0710.1568",
    archivePrefix = "arXiv",
    primaryClass = "hep-th",
    reportNumber = "ITP-UU-07-50, SPIN-07-37, TCDMATH-07-15",
    doi = "10.1088/1126-6708/2007/12/024",
    journal = "JHEP",
    volume = "12",
    pages = "024",
    year = "2007"
}

@article{Bombardelli:2009ns,
    author = "Bombardelli, Diego and Fioravanti, Davide and Tateo, Roberto",
    title = "{Thermodynamic Bethe Ansatz for planar AdS/CFT: A Proposal}",
    eprint = "0902.3930",
    archivePrefix = "arXiv",
    primaryClass = "hep-th",
    doi = "10.1088/1751-8113/42/37/375401",
    journal = "J. Phys. A",
    volume = "42",
    pages = "375401",
    year = "2009"
}

@article{Arutyunov:2009ur,
    author = "Arutyunov, Gleb and Frolov, Sergey",
    title = "{Thermodynamic Bethe Ansatz for the AdS$_5$ $\times$ S$_5$ Mirror Model}",
    eprint = "0903.0141",
    archivePrefix = "arXiv",
    primaryClass = "hep-th",
    reportNumber = "ITP-UU-09-09, SPIN-09-09, TCDMATH-09-09, HMI-09-05",
    doi = "10.1088/1126-6708/2009/05/068",
    journal = "JHEP",
    volume = "05",
    pages = "068",
    year = "2009"
}

@article{Gromov:2009bc,
    author = "Gromov, Nikolay and Kazakov, Vladimir and Kozak, Andrii and Vieira, Pedro",
    title = "{Exact Spectrum of Anomalous Dimensions of Planar N = 4 Supersymmetric Yang-Mills Theory: TBA and excited states}",
    eprint = "0902.4458",
    archivePrefix = "arXiv",
    primaryClass = "hep-th",
    reportNumber = "DESY-09-041",
    doi = "10.1007/s11005-010-0374-8",
    journal = "Lett. Math. Phys.",
    volume = "91",
    pages = "265--287",
    year = "2010"
}

@article{Gromov:2013pga,
    author = "Gromov, Nikolay and Kazakov, Vladimir and Leurent, Sebastien and Volin, Dmytro",
    title = "{Quantum Spectral Curve for Planar N = 4 Super-Yang-Mills Theory}",
    eprint = "1305.1939",
    archivePrefix = "arXiv",
    primaryClass = "hep-th",
    reportNumber = "IMPERIAL-TP-13-SL-02",
    doi = "10.1103/PhysRevLett.112.011602",
    journal = "Phys. Rev. Lett.",
    volume = "112",
    number = "1",
    pages = "011602",
    year = "2014"
}

@article{Gromov:2014caa,
    author = "Gromov, Nikolay and Kazakov, Vladimir and Leurent, Sébastien and Volin, Dmytro",
    title = "{Quantum spectral curve for arbitrary state/operator in AdS$_{5}$/CFT$_{4}$}",
    eprint = "1405.4857",
    archivePrefix = "arXiv",
    primaryClass = "hep-th",
    reportNumber = "NORDITA-2014-60-TCDMATH-14-05",
    doi = "10.1007/JHEP09(2015)187",
    journal = "JHEP",
    volume = "09",
    pages = "187",
    year = "2015"
}

@article{Gromov:2017blm,
    author = "Gromov, Nikolay",
    title = "{Introduction to the Spectrum of N = 4 SYM and the Quantum Spectral Curve}",
    eprint = "1708.03648",
    archivePrefix = "arXiv",
    primaryClass = "hep-th",
    month = "8",
    year = "2017"
}

@article{Levkovich-Maslyuk:2019awk,
    author = "Levkovich-Maslyuk, Fedor",
    title = "{A review of the AdS/CFT Quantum Spectral Curve}",
    eprint = "1911.13065",
    archivePrefix = "arXiv",
    primaryClass = "hep-th",
    doi = "10.1088/1751-8121/ab7137",
    journal = "J. Phys. A",
    volume = "53",
    number = "28",
    pages = "283004",
    year = "2020"
}

@article{Beisert:2005tm,
    author = "Beisert, Niklas",
    title = "{The SU(2$/$2) dynamic S-matrix}",
    eprint = "hep-th/0511082",
    archivePrefix = "arXiv",
    reportNumber = "PUTP-2181, NSF-KITP-05-92",
    doi = "10.4310/ATMP.2008.v12.n5.a1",
    journal = "Adv. Theor. Math. Phys.",
    volume = "12",
    pages = "945--979",
    year = "2008"
}

@article{Dorey:2006dq,
    author = "Dorey, Nick",
    title = "{Magnon Bound States and the AdS/CFT Correspondence}",
    eprint = "hep-th/0604175",
    archivePrefix = "arXiv",
    doi = "10.1088/0305-4470/39/41/S18",
    journal = "J. Phys. A",
    volume = "39",
    pages = "13119--13128",
    year = "2006"
}

@article{Chen:2006gq,
    author = "Chen, Heng-Yu and Dorey, Nick and Okamura, Keisuke",
    title = "{On the scattering of magnon boundstates}",
    eprint = "hep-th/0608047",
    archivePrefix = "arXiv",
    reportNumber = "DAMTP-06-62, UT-06-16",
    doi = "10.1088/1126-6708/2006/11/035",
    journal = "JHEP",
    volume = "11",
    pages = "035",
    year = "2006"
}

@article{deLeeuw:2008dp,
    author = "de Leeuw, Marius",
    title = "{Bound States, Yangian Symmetry and Classical r-matrix for the AdS$_5$ $\times$ S$^5$ Superstring}",
    eprint = "0804.1047",
    archivePrefix = "arXiv",
    primaryClass = "hep-th",
    reportNumber = "ITP-UU-08-18, SPIN-08-17",
    doi = "10.1088/1126-6708/2008/06/085",
    journal = "JHEP",
    volume = "06",
    pages = "085",
    year = "2008"
}

@article{Arutyunov:2008zt,
    author = "Arutyunov, Gleb and Frolov, Sergey",
    title = "{The S-matrix of String Bound States}",
    eprint = "0803.4323",
    archivePrefix = "arXiv",
    primaryClass = "hep-th",
    reportNumber = "ITP-UU-08-15, SPIN-08-14, TCDMATH-08-03",
    doi = "10.1016/j.nuclphysb.2008.06.005",
    journal = "Nucl. Phys. B",
    volume = "804",
    pages = "90--143",
    year = "2008"
}

@article{Arutyunov:2009mi,
    author = "Arutyunov, Gleb and de Leeuw, Marius and Torrielli, Alessandro",
    title = "{The Bound State S-Matrix for AdS$_5$ $\times$ S$^5$ Superstring}",
    eprint = "0902.0183",
    archivePrefix = "arXiv",
    primaryClass = "hep-th",
    reportNumber = "ITP-UU-09-06, SPIN-09-06",
    doi = "10.1016/j.nuclphysb.2009.03.024",
    journal = "Nucl. Phys. B",
    volume = "819",
    pages = "319--350",
    year = "2009"
}

@article{Chen:2006gp,
    author = "Chen, Heng-Yu and Dorey, Nick and Okamura, Keisuke",
    title = "{The Asymptotic spectrum of the N = 4 super Yang-Mills spin chain}",
    eprint = "hep-th/0610295",
    archivePrefix = "arXiv",
    reportNumber = "DAMTP-06-64, UY-06-17",
    doi = "10.1088/1126-6708/2007/03/005",
    journal = "JHEP",
    volume = "03",
    pages = "005",
    year = "2007"
}

@article{Matsumoto:2014cka,
    author = "Matsumoto, Takuya and Molev, Alexander",
    title = "{Representations of centrally extended Lie superalgebra psl(2$/$2)}",
    eprint = "1405.3420",
    archivePrefix = "arXiv",
    primaryClass = "math.RT",
    doi = "10.1063/1.4896396",
    journal = "J. Math. Phys.",
    volume = "55",
    pages = "091704",
    year = "2014"
}

@article{Berenstein:2002jq,
    author = "Berenstein, David Eliecer and Maldacena, Juan Martin and Nastase, Horatiu Stefan",
    title = "{Strings in flat space and pp waves from N = 4 superYang-Mills}",
    eprint = "hep-th/0202021",
    archivePrefix = "arXiv",
    doi = "10.1088/1126-6708/2002/04/013",
    journal = "JHEP",
    volume = "04",
    pages = "013",
    year = "2002"
}

@article{Beisert:2003yb,
    author = "Beisert, Niklas and Staudacher, Matthias",
    title = "{The N = 4 SYM integrable super spin chain}",
    eprint = "hep-th/0307042",
    archivePrefix = "arXiv",
    reportNumber = "AEI-2003-057",
    doi = "10.1016/j.nuclphysb.2003.08.015",
    journal = "Nucl. Phys. B",
    volume = "670",
    pages = "439--463",
    year = "2003"
}

@article{Beisert:2003tq,
    author = "Beisert, N. and Kristjansen, C. and Staudacher, M.",
    title = "{The Dilatation operator of conformal N = 4 superYang-Mills theory}",
    eprint = "hep-th/0303060",
    archivePrefix = "arXiv",
    reportNumber = "AEI-2003-028",
    doi = "10.1016/S0550-3213(03)00406-1",
    journal = "Nucl. Phys. B",
    volume = "664",
    pages = "131--184",
    year = "2003"
}

@article{Metsaev:1998it,
    author = "Metsaev, R. R. and Tseytlin, Arkady A.",
    title = "{Type IIB superstring action in AdS$_5$ $\times$ S$^5$ background}",
    eprint = "hep-th/9805028",
    archivePrefix = "arXiv",
    reportNumber = "FIAN-TD-98-21, IMPERIAL-TP-97-98-44, NSF-ITP-98-055",
    doi = "10.1016/S0550-3213(98)00570-7",
    journal = "Nucl. Phys. B",
    volume = "533",
    pages = "109--126",
    year = "1998"
}

@article{Brink:1976bc,
    author = "Brink, Lars and Schwarz, John H. and Scherk, Joel",
    title = "{Supersymmetric Yang-Mills Theories}",
    reportNumber = "CALT-68-574",
    doi = "10.1016/0550-3213(77)90328-5",
    journal = "Nucl. Phys. B",
    volume = "121",
    pages = "77--92",
    year = "1977"
}

@article{Arutyunov:2009ga,
    author = "Arutyunov, Gleb and Frolov, Sergey",
    title = "{Foundations of the AdS$_{5} \times S^{5}$ Superstring. Part I}",
    eprint = "0901.4937",
    archivePrefix = "arXiv",
    primaryClass = "hep-th",
    reportNumber = "ITP-UU-09-05, SPIN-09-05, TCD-MATH-09-06, HMI-09-03",
    doi = "10.1088/1751-8113/42/25/254003",
    journal = "J. Phys. A",
    volume = "42",
    pages = "254003",
    year = "2009"
}

@article{Bena:2003wd,
    author = "Bena, Iosif and Polchinski, Joseph and Roiban, Radu",
    title = "{Hidden symmetries of the AdS$_5$ $\times$ S$^5$ superstring}",
    eprint = "hep-th/0305116",
    archivePrefix = "arXiv",
    reportNumber = "NSF-KITP-03-34, UCLA-03-TEP-14",
    doi = "10.1103/PhysRevD.69.046002",
    journal = "Phys. Rev. D",
    volume = "69",
    pages = "046002",
    year = "2004"
}

@article{Frolov:2006cc,
    author = "Frolov, Sergey and Plefka, Jan and Zamaklar, Marija",
    title = "{The AdS$_5$ $\times$ S$^5$ superstring in light-cone gauge and its Bethe equations}",
    eprint = "hep-th/0603008",
    archivePrefix = "arXiv",
    reportNumber = "AEI-2006-011, HU-EP-06-08",
    doi = "10.1088/0305-4470/39/41/S15",
    journal = "J. Phys. A",
    volume = "39",
    pages = "13037--13082",
    year = "2006"
}

@article{Arutyunov:2004yx,
    author = "Arutyunov, Gleb and Frolov, Sergey",
    title = "{Integrable Hamiltonian for classical strings on AdS$_5$ $\times$ S$^5$}",
    eprint = "hep-th/0411089",
    archivePrefix = "arXiv",
    reportNumber = "AEI-2004-105",
    doi = "10.1088/1126-6708/2005/02/059",
    journal = "JHEP",
    volume = "02",
    pages = "059",
    year = "2005"
}

@article{Arutyunov:2006ak,
    author = "Arutyunov, Gleb and Frolov, Sergey and Plefka, Jan and Zamaklar, Marija",
    title = "{The Off-shell Symmetry Algebra of the Light-cone AdS$_5$ $\times$ S$^5$ Superstring}",
    eprint = "hep-th/0609157",
    archivePrefix = "arXiv",
    reportNumber = "AEI-2006-071, HU-EP-06-31, ITP-UU-06-39, SPIN-06-33, TCDMATH-06-13",
    doi = "10.1088/1751-8113/40/13/018",
    journal = "J. Phys. A",
    volume = "40",
    pages = "3583--3606",
    year = "2007"
}

@article{Arutyunov:2004vx,
    author = "Arutyunov, Gleb and Frolov, Sergey and Staudacher, Matthias",
    title = "{Bethe ansatz for quantum strings}",
    eprint = "hep-th/0406256",
    archivePrefix = "arXiv",
    reportNumber = "AEI-2004-046",
    doi = "10.1088/1126-6708/2004/10/016",
    journal = "JHEP",
    volume = "10",
    pages = "016",
    year = "2004"
}

@article{Klose:2006zd,
    author = "Klose, Thomas and McLoughlin, Tristan and Roiban, Radu and Zarembo, Konstantin",
    title = "{Worldsheet scattering in AdS$_5$ $\times$ S$^5$}",
    eprint = "hep-th/0611169",
    archivePrefix = "arXiv",
    reportNumber = "ITEP-TH-61-06, UUITP-15-06",
    doi = "10.1088/1126-6708/2007/03/094",
    journal = "JHEP",
    volume = "03",
    pages = "094",
    year = "2007"
}

@article{Roiban:2007jf,
    author = "Roiban, R. and Tirziu, A. and Tseytlin, Arkady A.",
    title = "{Two-loop world-sheet corrections in AdS$_5$ $\times$ S$^5$ superstring}",
    eprint = "0704.3638",
    archivePrefix = "arXiv",
    primaryClass = "hep-th",
    reportNumber = "IMPERIAL-TP-AT-2007-1",
    doi = "10.1088/1126-6708/2007/07/056",
    journal = "JHEP",
    volume = "07",
    pages = "056",
    year = "2007"
}

@article{Klose:2007rz,
    author = "Klose, T. and McLoughlin, T. and Minahan, J. A. and Zarembo, K.",
    title = "{World-sheet scattering in AdS$_5$ $\times$ S$^5$ at two loops}",
    eprint = "0704.3891",
    archivePrefix = "arXiv",
    primaryClass = "hep-th",
    reportNumber = "ITEP-TH-17-7, UUITP-07-07",
    doi = "10.1088/1126-6708/2007/08/051",
    journal = "JHEP",
    volume = "08",
    pages = "051",
    year = "2007"
}

@article{Janik:2006dc,
    author = "Janik, Romuald A.",
    title = "{The AdS$_5$ $\times$ S$^5$ superstring worldsheet S-matrix and crossing symmetry}",
    eprint = "hep-th/0603038",
    archivePrefix = "arXiv",
    doi = "10.1103/PhysRevD.73.086006",
    journal = "Phys. Rev. D",
    volume = "73",
    pages = "086006",
    year = "2006"
}

@article{Hernandez:2006tk,
    author = "Hernández, Rafael and López, Esperanza",
    title = "{Quantum corrections to the string Bethe ansatz}",
    eprint = "hep-th/0603204",
    archivePrefix = "arXiv",
    reportNumber = "CERN-PH-TH-2006-048, IFT-UAM-CSIC-06-14",
    doi = "10.1088/1126-6708/2006/07/004",
    journal = "JHEP",
    volume = "07",
    pages = "004",
    year = "2006"
}

@article{Arutyunov:2006iu,
    author = "Arutyunov, G. and Frolov, S.",
    title = "{On AdS$_5$ $\times$ S$^5$ String S-matrix}",
    eprint = "hep-th/0604043",
    archivePrefix = "arXiv",
    reportNumber = "ITP-UU-06-15, SPIN-06-13",
    doi = "10.1016/j.physletb.2006.06.064",
    journal = "Phys. Lett. B",
    volume = "639",
    pages = "378--382",
    year = "2006"
}

@article{Beisert:2006ib,
    author = "Beisert, Niklas and Hernández, Rafael and López, Esperanza",
    title = "{A Crossing-symmetric phase for AdS$_5$ $\times$ S$^5$ strings}",
    eprint = "hep-th/0609044",
    archivePrefix = "arXiv",
    reportNumber = "AEI-2006-068, CERN-PH-TH-2006-176, IFT-UAM-CSIC-06-44, PUTP-2208",
    doi = "10.1088/1126-6708/2006/11/070",
    journal = "JHEP",
    volume = "11",
    pages = "070",
    year = "2006"
}

@article{Beisert:2006ez,
    author = "Beisert, Niklas and Eden, Burkhard and Staudacher, Matthias",
    title = "{Transcendentality and Crossing}",
    eprint = "hep-th/0610251",
    archivePrefix = "arXiv",
    reportNumber = "AEI-2006-079, ITP-UU-06-44, SPIN-06-34",
    doi = "10.1088/1742-5468/2007/01/P01021",
    journal = "J. Stat. Mech.",
    volume = "0701",
    pages = "P01021",
    year = "2007"
}

@article{Dorey:2007xn,
    author = "Dorey, Nick and Hofman, Diego M. and Maldacena, Juan Martin",
    title = "{On the Singularities of the Magnon S-matrix}",
    eprint = "hep-th/0703104",
    archivePrefix = "arXiv",
    doi = "10.1103/PhysRevD.76.025011",
    journal = "Phys. Rev. D",
    volume = "76",
    pages = "025011",
    year = "2007"
}

@article{Spill:2008tp,
    author = "Spill, Fabian and Torrielli, Alessandro",
    title = "{On Drinfeld's second realization of the AdS/CFT su(2$/$2) Yangian}",
    eprint = "0803.3194",
    archivePrefix = "arXiv",
    primaryClass = "hep-th",
    reportNumber = "MIT-CTP-3935, IMPERIAL-TP-08-FS-01, HU-EP-08-05",
    doi = "10.1016/j.geomphys.2009.01.001",
    journal = "J. Geom. Phys.",
    volume = "59",
    pages = "489--502",
    year = "2009"
}

@article{Beisert:2014hya,
    author = "Beisert, Niklas and de Leeuw, Marius",
    title = "{The RTT realization for the deformed gl(2$/$2) Yangian}",
    eprint = "1401.7691",
    archivePrefix = "arXiv",
    primaryClass = "math-ph",
    doi = "10.1088/1751-8113/47/30/305201",
    journal = "J. Phys. A",
    volume = "47",
    pages = "305201",
    year = "2014"
}

@article{Beisert:2016qei,
    author = "Beisert, Niklas and de Leeuw, Marius and Hecht, Reimar",
    title = "{Maximally extended sl(2$/$2) as a quantum double}",
    eprint = "1602.04988",
    archivePrefix = "arXiv",
    primaryClass = "math-ph",
    doi = "10.1088/1751-8113/49/43/434005",
    journal = "J. Phys. A",
    volume = "49",
    number = "43",
    pages = "434005",
    year = "2016"
}

@article{Matsumoto:2022nrk,
    author = "Matsumoto, Takuya",
    title = "{Drinfeld realization of the centrally extended psl(2$/$2) Yangian algebra with the manifest coproducts}",
    eprint = "2208.11889",
    archivePrefix = "arXiv",
    primaryClass = "math.QA",
    doi = "10.1063/5.0124333",
    journal = "J. Math. Phys.",
    volume = "64",
    number = "4",
    pages = "041704",
    year = "2023"
}

@article{Gomez:2006va,
    author = "Gómez, César and Hernández, Rafael",
    title = "{The Magnon kinematics of the AdS/CFT correspondence}",
    eprint = "hep-th/0608029",
    archivePrefix = "arXiv",
    reportNumber = "CERN-PH-TH-2006-140, IFT-UAM-CSIC-06-37",
    doi = "10.1088/1126-6708/2006/11/021",
    journal = "JHEP",
    volume = "11",
    pages = "021",
    year = "2006"
}

@article{Plefka:2006ze,
    author = "Plefka, Jan and Spill, Fabian and Torrielli, Alessandro",
    title = "{On the Hopf algebra structure of the AdS/CFT S-matrix}",
    eprint = "hep-th/0608038",
    archivePrefix = "arXiv",
    reportNumber = "HU-EP-06-22",
    doi = "10.1103/PhysRevD.74.066008",
    journal = "Phys. Rev. D",
    volume = "74",
    pages = "066008",
    year = "2006"
}

@article{Beisert:2006fmy,
    author = "Beisert, Niklas",
    editor = "Faddeev, L. and Henneaux, M. and Kashaev, R. and Volkov, A. and Lambert, F.",
    title = "{The S-matrix of AdS/CFT and Yangian symmetry}",
    eprint = "0704.0400",
    archivePrefix = "arXiv",
    primaryClass = "nlin.SI",
    reportNumber = "AEI-2007-019",
    doi = "10.22323/1.038.0002",
    journal = "PoS",
    volume = "SOLVAY",
    pages = "002",
    year = "2006"
}

@article{Drummond:2009fd,
    author = "Drummond, James M. and Henn, Johannes M. and Plefka, Jan",
    editor = "Liu, Feng and Xiao, Zhigang and Zhuang, Pengfei",
    title = "{Yangian symmetry of scattering amplitudes in N = 4 super Yang-Mills theory}",
    eprint = "0902.2987",
    archivePrefix = "arXiv",
    primaryClass = "hep-th",
    reportNumber = "HU-EP-09-06, LAPTH-1308-09",
    doi = "10.1088/1126-6708/2009/05/046",
    journal = "JHEP",
    volume = "05",
    pages = "046",
    year = "2009"
}

@article{Beisert:2017pnr,
    author = "Beisert, Niklas and Garus, Aleksander and Rosso, Matteo",
    title = "{Yangian Symmetry and Integrability of Planar N = 4 Supersymmetric Yang-Mills Theory}",
    eprint = "1701.09162",
    archivePrefix = "arXiv",
    primaryClass = "hep-th",
    reportNumber = "HU-EP-17-03",
    doi = "10.1103/PhysRevLett.118.141603",
    journal = "Phys. Rev. Lett.",
    volume = "118",
    number = "14",
    pages = "141603",
    year = "2017"
}

@article{Matsumoto:2007rh,
    author = "Matsumoto, Takuya and Moriyama, Sanefumi and Torrielli, Alessandro",
    title = "{A Secret Symmetry of the AdS/CFT S-matrix}",
    eprint = "0708.1285",
    archivePrefix = "arXiv",
    primaryClass = "hep-th",
    reportNumber = "MIT-CTP-3853",
    doi = "10.1088/1126-6708/2007/09/099",
    journal = "JHEP",
    volume = "09",
    pages = "099",
    year = "2007"
}

@article{deLeeuw:2012jf,
    author = "de Leeuw, Marius and Matsumoto, Takuya and Moriyama, Sanefumi and Regelskis, Vidas and Torrielli, Alessandro",
    title = "{Secret Symmetries in AdS/CFT}",
    eprint = "1204.2366",
    archivePrefix = "arXiv",
    primaryClass = "hep-th",
    reportNumber = "DMUS-MP-12-04, NORDITA-2012-26",
    doi = "10.1088/0031-8949/86/02/028502",
    journal = "Phys. Scripta",
    volume = "02",
    pages = "028502",
    year = "2012"
}

@article{Moriyama:2007jt,
    author = "Moriyama, Sanefumi and Torrielli, Alessandro",
    title = "{A Yangian double for the AdS/CFT classical r-matrix}",
    eprint = "0706.0884",
    archivePrefix = "arXiv",
    primaryClass = "hep-th",
    reportNumber = "MIT-CTP-3843",
    doi = "10.1088/1126-6708/2007/06/083",
    journal = "JHEP",
    volume = "06",
    pages = "083",
    year = "2007"
}

@article{Torrielli:2007mc,
    author = "Torrielli, Alessandro",
    title = "{Classical r-matrix of the su(2$/$2) SYM spin-chain}",
    eprint = "hep-th/0701281",
    archivePrefix = "arXiv",
    reportNumber = "MIT-CTP-3809",
    doi = "10.1103/PhysRevD.75.105020",
    journal = "Phys. Rev. D",
    volume = "75",
    pages = "105020",
    year = "2007"
}

@article{Beisert:2017xqx,
    author = "Beisert, Niklas and Hecht, Reimar and Hoare, Ben",
    title = "{Maximally extended sl(2$/$2), q-deformed d(2,1;$\epsilon$) and 3D kappa-Poincaré}",
    eprint = "1704.05093",
    archivePrefix = "arXiv",
    primaryClass = "math-ph",
    doi = "10.1088/1751-8121/aa7a2f",
    journal = "J. Phys. A",
    volume = "50",
    number = "31",
    pages = "314003",
    year = "2017"
}

@article{Matsumoto:2008ww,
    author = "Matsumoto, Takuya and Moriyama, Sanefumi",
    title = "{An Exceptional Algebraic Origin of the AdS/CFT Yangian Symmetry}",
    eprint = "0803.1212",
    archivePrefix = "arXiv",
    primaryClass = "hep-th",
    doi = "10.1088/1126-6708/2008/04/022",
    journal = "JHEP",
    volume = "04",
    pages = "022",
    year = "2008"
}

@article{Delduc:2013qra,
    author = "Delduc, Francois and Magro, Marc and Vicedo, Benoit",
    title = "{An integrable deformation of the AdS$_5$ $\times$ S$^5$ superstring action}",
    eprint = "1309.5850",
    archivePrefix = "arXiv",
    primaryClass = "hep-th",
    doi = "10.1103/PhysRevLett.112.051601",
    journal = "Phys. Rev. Lett.",
    volume = "112",
    number = "5",
    pages = "051601",
    year = "2014"
}

@article{Arutyunov:2013ega,
    author = "Arutyunov, Gleb and Borsato, Riccardo and Frolov, Sergey",
    title = "{S-matrix for strings on $\eta$-deformed AdS$_5$ $\times$ S$^5$}",
    eprint = "1312.3542",
    archivePrefix = "arXiv",
    primaryClass = "hep-th",
    reportNumber = "ITP-UU-13-31, SPIN-13-23, HU-MATHEMATIK-2013-24, TCD-MATH-13-16",
    doi = "10.1007/JHEP04(2014)002",
    journal = "JHEP",
    volume = "04",
    pages = "002",
    year = "2014"
}

@article{Delduc:2014kha,
    author = "Delduc, Francois and Magro, Marc and Vicedo, Benoit",
    title = "{Derivation of the action and symmetries of the q-deformed AdS$_5$ $\times$ S$^5$ superstring}",
    eprint = "1406.6286",
    archivePrefix = "arXiv",
    primaryClass = "hep-th",
    doi = "10.1007/JHEP10(2014)132",
    journal = "JHEP",
    volume = "10",
    pages = "132",
    year = "2014"
}

@article{Nahm:1977tg,
    author = "Nahm, W.",
    title = "{Supersymmetries and their Representations}",
    reportNumber = "CERN-TH-2341",
    doi = "10.1016/0550-3213(78)90218-3",
    journal = "Nucl. Phys. B",
    volume = "135",
    pages = "149",
    year = "1978"
}

@article{VanderJeugt:1985,
    author = {Van der Jeugt,J. },
    title = {Irreducible representations of the exceptional Lie superalgebras D(2,1;$\alpha$)},
    journal = {J. Math. Phys.},
    volume = {26},
    number = {5},
    pages = {913-924},
    year = {1985},
    doi = {10.1063/1.526547}
}

@article{deLeeuw:2011fr,
    author = "de Leeuw, Marius and Regelskis, Vidas and Torrielli, Alessandro",
    title = "{The Quantum Affine Origin of the AdS/CFT Secret Symmetry}",
    eprint = "1112.4989",
    archivePrefix = "arXiv",
    primaryClass = "hep-th",
    reportNumber = "DMUS-MP-11-02",
    doi = "10.1088/1751-8113/45/17/175202",
    journal = "J. Phys. A",
    volume = "45",
    pages = "175202",
    year = "2012"
}

@article{Beisert:2008tw,
    author = "Beisert, Niklas and Koroteev, Peter",
    title = "{Quantum Deformations of the One-Dimensional Hubbard Model}",
    eprint = "0802.0777",
    archivePrefix = "arXiv",
    primaryClass = "hep-th",
    reportNumber = "AEI-2008-003, ITEP-TH-06-08",
    doi = "10.1088/1751-8113/41/25/255204",
    journal = "J. Phys. A",
    volume = "41",
    pages = "255204",
    year = "2008"
}

@article{Beisert:2011wq,
    author = "Beisert, Niklas and Galleas, Wellington and Matsumoto, Takuya",
    title = "{A Quantum Affine Algebra for the Deformed Hubbard Chain}",
    eprint = "1102.5700",
    archivePrefix = "arXiv",
    primaryClass = "math-ph",
    reportNumber = "AEI-2011-005",
    doi = "10.1088/1751-8113/45/36/365206",
    journal = "J. Phys. A",
    volume = "45",
    pages = "365206",
    year = "2012"
}

@article{Seibold:2020ywq,
    author = "Seibold, Fiona K. and Van Tongeren, Stijn J. and Zimmermann, Yannik",
    title = "{The twisted story of worldsheet scattering in $\eta$-deformed AdS$_5$ $\times$ S$^5$}",
    eprint = "2007.09136",
    archivePrefix = "arXiv",
    primaryClass = "hep-th",
    doi = "10.1007/JHEP12(2020)043",
    journal = "JHEP",
    volume = "12",
    pages = "043",
    year = "2020"
}

@article{Beisert:2014qba,
    author = "Beisert, Niklas and Broedel, Johannes and Rosso, Matteo",
    title = "{On Yangian-invariant regularization of deformed on-shell diagrams in N = 4 super-Yang-Mills theory}",
    eprint = "1401.7274",
    archivePrefix = "arXiv",
    primaryClass = "hep-th",
    doi = "10.1088/1751-8113/47/36/365402",
    journal = "J. Phys. A",
    volume = "47",
    pages = "365402",
    year = "2014"
}

@article{Rej:2005qt,
    author = "Rej, Adam and Serban, Didina and Staudacher, Matthias",
    title = "{Planar N = 4 gauge theory and the Hubbard model}",
    eprint = "hep-th/0512077",
    archivePrefix = "arXiv",
    reportNumber = "AEI-2005-164, SPHT-T05-190, NSF-KITP-05-84",
    doi = "10.1088/1126-6708/2006/03/018",
    journal = "JHEP",
    volume = "03",
    pages = "018",
    year = "2006"
}

@article {Tolstoy:1992,
    author = "Tolstoy, V. N. and Khoroshkin, S. M.", 
    title = {Universal {$R$}-matrix for quantized nontwisted affine {L}ie algebras},
    journal = {Funktsional'nyi Analiz i ego Prilozheniya},
    volume = {26},
    year = {1992},
    number = {1},
    pages = {85--88},
    issn = {0374-1990,2305-2899},
    doi = {10.1007/BF01077085},
}

@article{Khoroshkin:1991,
author = {S. M. Khoroshkin and V. N. Tolstoy},
title = {Universal $R$-matrix for quantized (super)algebras},
volume = {141},
journal = {Communications in Mathematical Physics},
number = {3},
publisher = {Springer},
pages = {599 -- 617},
year = {1991},
}

@article{Heckenberger:2007ry,
    author = "Heckenberger, Istvan and Spill, Fabian and Torrielli, Alessandro and Yamane, Hiroyuki",
    title = "{Drinfeld second realization of the quantum affine superalgebras of D(1)(2,1;x) via the Weyl groupoid}",
    eprint = "0705.1071",
    archivePrefix = "arXiv",
    primaryClass = "math.QA",
    reportNumber = "MIT-CTP-3835, HU-EP-07-15",
    journal = "Publ. Res. Inst. Math. Sci. Kyoto B",
    volume = "8",
    pages = "171",
    year = "2008"
}

@article{Beisert:2018zxs,
    author = "Beisert, Niklas and Garus, Aleksander and Rosso, Matteo",
    title = "{Yangian Symmetry for the Action of Planar N = 4 Super Yang-Mills and N = 6 Super Chern-Simons Theories}",
    eprint = "1803.06310",
    archivePrefix = "arXiv",
    primaryClass = "hep-th",
    reportNumber = "HU-EP-18/08, HU-EP-18-08",
    doi = "10.1103/PhysRevD.98.046006",
    journal = "Phys. Rev. D",
    volume = "98",
    number = "4",
    pages = "046006",
    year = "2018"
}

@inproceedings{Dolan:2004ps,
    author = "Dolan, Louise and Nappi, Chiara R. and Witten, Edward",
    title = "{Yangian symmetry in D = 4 superconformal Yang-Mills theory}",
    booktitle = "{3rd International Symposium on Quantum Theory and Symmetries}",
    eprint = "hep-th/0401243",
    archivePrefix = "arXiv",
    doi = "10.1142/9789812702340_0036",
    pages = "300--315",
    month = "1",
    year = "2004"
}

\end{bibtex}
\fi

\bibliographystyle{nb}
\bibliography{\jobname}

\end{document}